\documentclass[twocolumn,prl,superscriptaddress,notitlepage]{revtex4}

\usepackage{graphicx}
\usepackage{comment}
\usepackage{amsmath,amssymb,amsthm} 
\usepackage{verbatim}
\usepackage{float}
\setcitestyle{super}

\usepackage[sort&compress]{natbib}

\begin{document}

\title{Predicting how nanoconfinement changes the relaxation time of a supercooled liquid}
\author{Trond S. Ingebrigtsen}
\email{trond@ruc.dk}
\affiliation{DNRF Centre ``Glass and Time'', IMFUFA, Department of Sciences, Roskilde University, Postbox 260, DK-4000 Roskilde, Denmark}
\author{Jeffrey R. Errington}
\affiliation{Department of Chemical and Biological Engineering, University at Buffalo, The State University of New York, Buffalo, New York 14260, USA}
\author{Thomas M. Truskett}
\affiliation{McKetta Department of Chemical Engineering and Institute for Theoretical Chemistry, University of Texas at Austin, Austin, Texas 78712, USA}
\author{Jeppe C. Dyre}
\affiliation{DNRF Centre ``Glass and Time'', IMFUFA, Department of Sciences, Roskilde University, Postbox 260, DK-4000 Roskilde, Denmark}

\date{\today}

\begin{abstract}
The properties of nanoconfined fluids can be strikingly different from those of bulk liquids. A basic unanswered question is whether the equilibrium and dynamic consequences of confinement are related to each other in a simple way. We study this question by simulation of a liquid comprising asymmetric dumbbell-shaped molecules, which can be deeply supercooled without crystallizing. We find that the dimensionless structural relaxation times $-$ spanning six decades as a function of temperature, density, and degree of confinement $-$ collapse when plotted versus excess entropy. The data also collapse when plotted versus excess isochoric heat capacity, a behaviour that follows from the existence of isomorphs in the bulk and confined states.
\end{abstract}

\maketitle

That confined liquids microscopically relax and flow with different characteristic time scales than bulk liquids is hardly surprising. Confining boundaries bias the spatial distribution of the constituent molecules and the ways by which those molecules can dynamically rearrange. These effects play important roles in the design of coating, nanopatterning, and nanomanufacturing technologies\cite{bhushan1995,whitesides2006}. As a result, they have already been experimentally characterized for a wide variety of material systems, including small-molecule fluids\cite{drake1990,granick1991,morineau2002,jackson1991,teboul2005,alcoutlabi2005,coasne2011,richert2011}, 
polymers\cite{keddie1994,sergei2006,forrest2001,ellison2003,rittigstein2007,paeng2012}, ionic liquids\cite{iacob2012}, liquid crystals\cite{ji2009}, and dense colloidal suspensions\cite{nugent2007,eral2009,michailidou2009,watanabe2011,edmond2012}, and studied extensively via molecular simulations\cite{fehr1995,torres2000,scheidler2000,starr2002,baljon2005,kurzidim2009,watanabe2011,starr2011,betancourt2013}.  
Recent reviews of confined-liquid behavior may be found in, e.g., Refs. \onlinecite{alba2006,richert2011}.

Unfortunately, successful theories for predicting the dynamics of inhomogeneous fluids have been slower to emerge. Here, we explore the possibility of a novel approach for predicting how confinement affects the dynamics of viscous fluids. The central idea is motivated by the observation from molecular simulations that, under equilibrium conditions, key dimensionless ``reduced'' quantities for confined fluids closely correspond to those of homogeneous bulk fluids with the same excess entropy\cite{hsconfined, LJconfinedsmoothwalls,goel2009,dumbbellconfinedroughwalls,borah2012} (relative to an ideal gas at the same density and temperature). The excess entropy can be computed using Monte Carlo methods\cite{dumbbellconfinedroughwalls} or predicted from classical density-functional theories\cite{gulati1997,goel2009}. An open question is whether this observed correspondence between dynamics and excess entropy applies for fluids in deeply supercooled liquid states approaching the glass transition, where highly nontrivial dynamic effects of confinement are observed. Another open question is whether thermodynamic properties other than the excess entropy can be used to predict the dynamics in confinement. 

To investigate these questions we study the behavior of a glass-former comprising asymmetric dumbbell-shaped molecules\cite{moleculeshidden}. This model is perhaps the simplest single-component system that avoids freezing upon cooling or compression in confinement, allowing for a systematic comparison of the properties of supercooled states in both bulk and confined geometries. The latter is modeled as a slit-pore, i.e., a sandwich geometry, using a 9-3 Lennard-Jones wall potential. The pore geometry is ideal for exploring the physics of confinement, which can be difficult to extract from experiments on porous materials that often have a complex distribution of pore sizes, geometries, and fluid-pore interactions\cite{schoeffel2012,coasne2011}. The possible effects of corrugation and realistic pore geometries and interactions on scaling behavior are discussed in Refs. \onlinecite{LJconfinedsmoothwalls,dumbbellconfinedroughwalls,krekelberg2011}. 

Molecular dynamics and Monte Carlo methods were used to simulate the model using high-speed graphics cards (GPUs, see http://rumd.org), obtaining for all state points consistent results from the two methods. Details of the model, simulations, units, etc. are provided in the Supplemental Material. The longest production run was four billion time steps (approximately 360 GPU hours), requiring more than two months of prior equilibration. The main results obtained are proof of the existence of isomorphs in a strongly inhomogeneous fluid and, as a consequence of this, in particular that the excess isochoric specific heat controls the relaxation time in the same sense as the excess entropy does. 

\begin{figure}[H]
  \centering
  \includegraphics[width=80mm]{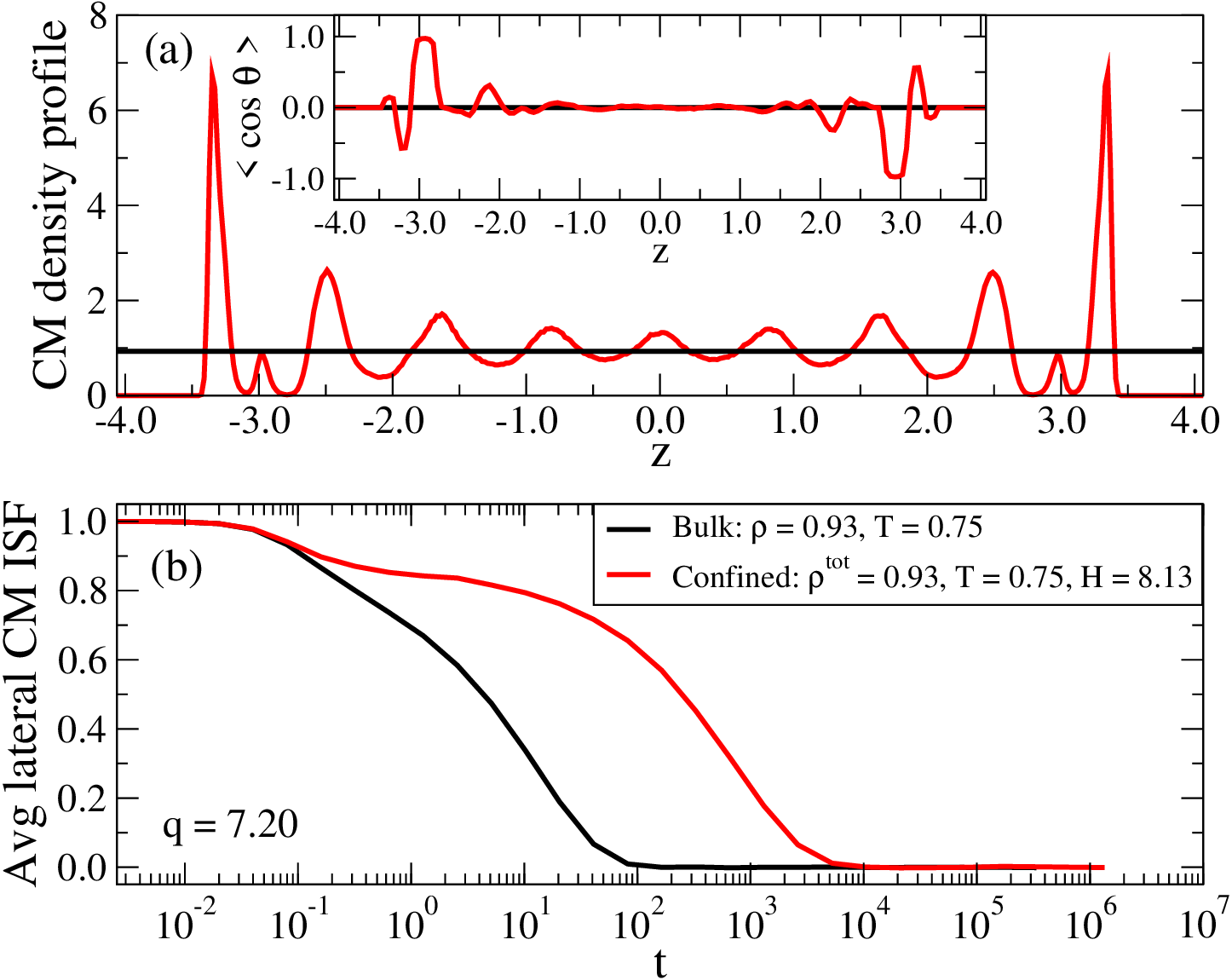}
  \caption{Structure and dynamics of the asymmetric dumbbell model for the nanoscale-confined (red) and the bulk (black) liquid at the same temperature and average density. (a) The molecular center-of-mass density profile in the direction normal to the confining slit-pore walls ($z$-direction); the inset shows the average orientation of the dumbbell molecules with respect to the $z$-axis (see the Supplemental Material). (b) The molecular center-of-mass incoherent intermediate scattering function for the wave vector $q=7.20$ parallel to the walls of the slit-pore.}
  \label{bulkconfinement}
\end{figure}

We begin the investigation by studying in Fig. \ref{bulkconfinement}(a) the structural changes induced by the confining slit-pore geometry. This figure shows in red the molecular center-of-mass density profile in the direction normal to the walls of the slit-pore ($z$-direction). There are significant density oscillations, particularly close to the walls. Probing the average orientation of the dumbbell molecules with respect to the $z$-axis (inset) reveals that preferred orientations emerge as the wall is approached. Both of these structural effects are absent in the bulk liquid, of course (black curves), and as shown in the Supplemental Material they lead to a heterogeneous dynamics that is substantially slower near the walls. Figure \ref{bulkconfinement}(b) shows the spatially averaged dynamics in confinement (red) and bulk liquid (black) at the same temperature and average density; it is two orders of magnitude slower under confinement than in the bulk liquid phase. In fact, two-step relaxation -- the hallmark of the supercooled viscous liquid state\cite{debenedetti2001} --  is seen in confinement but not at the corresponding bulk-liquid state point. The geometry thus has a pronounced effect on both structure and dynamics that cannot be accounted for by a trivial shifting or rescaling of the bulk data; this is observed in experimental realizations of similar systems\cite{eral2009,watanabe2011,edmond2012}.
\newline \newline

To investigate whether the reduction in entropy upon confinement predicts the slowing down Fig. \ref{excess} shows the reduced spatially averaged structural relaxation time $\tilde{\tau}_{\alpha}$ in bulk and confinement as a function of the excess entropy $S_{\rm ex}$ (the relaxation time is determined from the center-of-mass incoherent intermediate scattering function as described in detail in the Supplementary Material). 
Two different versions of the excess entropy is shown. One uses the total slit-pore volume, one corrects for the non-accessible volume close to the walls\cite{dumbbellconfinedroughwalls} (see the Supplemental Material). Both versions capture well the changes in the dynamics induced by confinement.
\newline \newline
\begin{figure}[H]
  \centering
  \includegraphics[width=80mm]{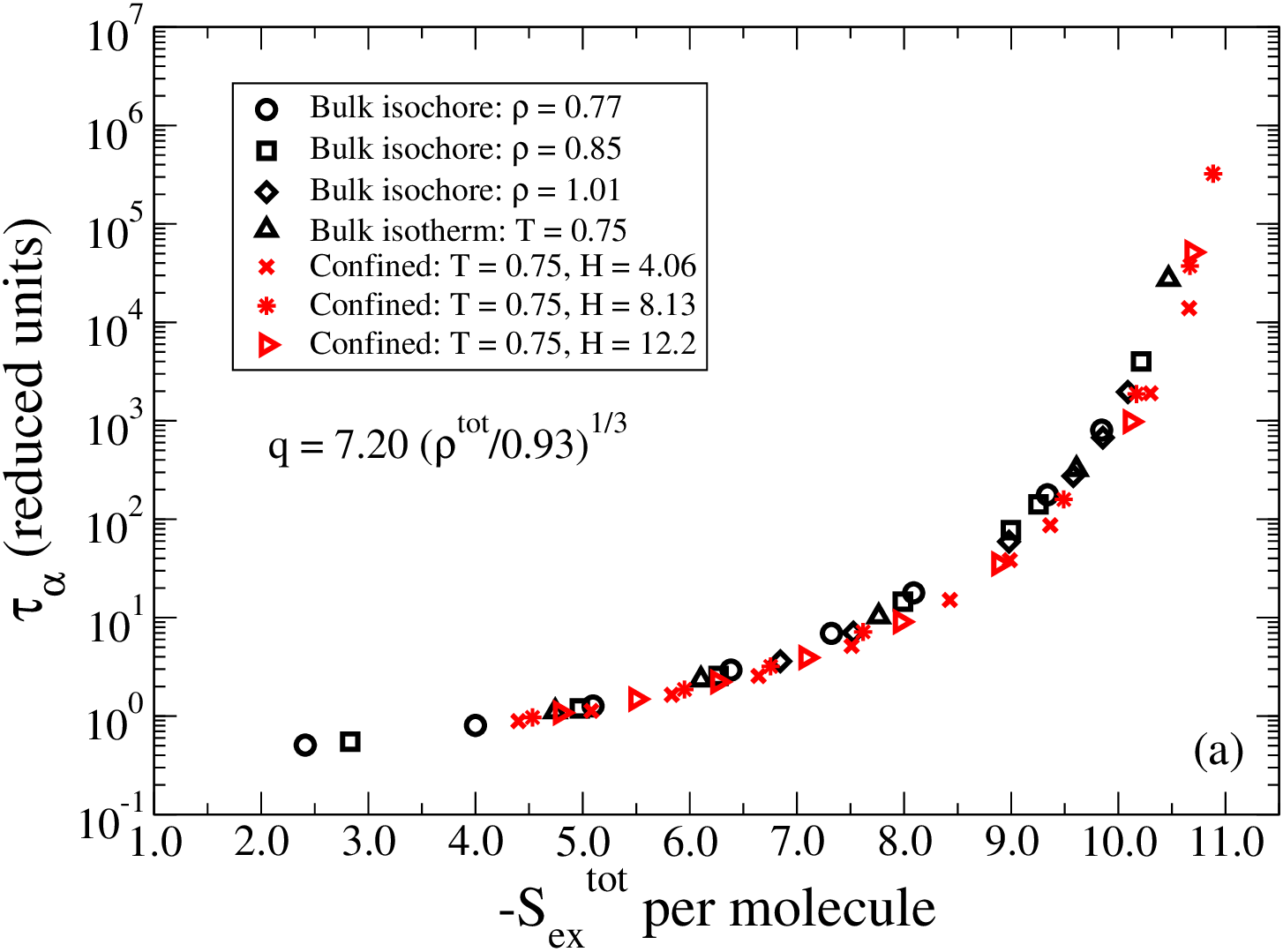}
\end{figure}
\begin{figure}[H]
  \centering
  \includegraphics[width=80mm]{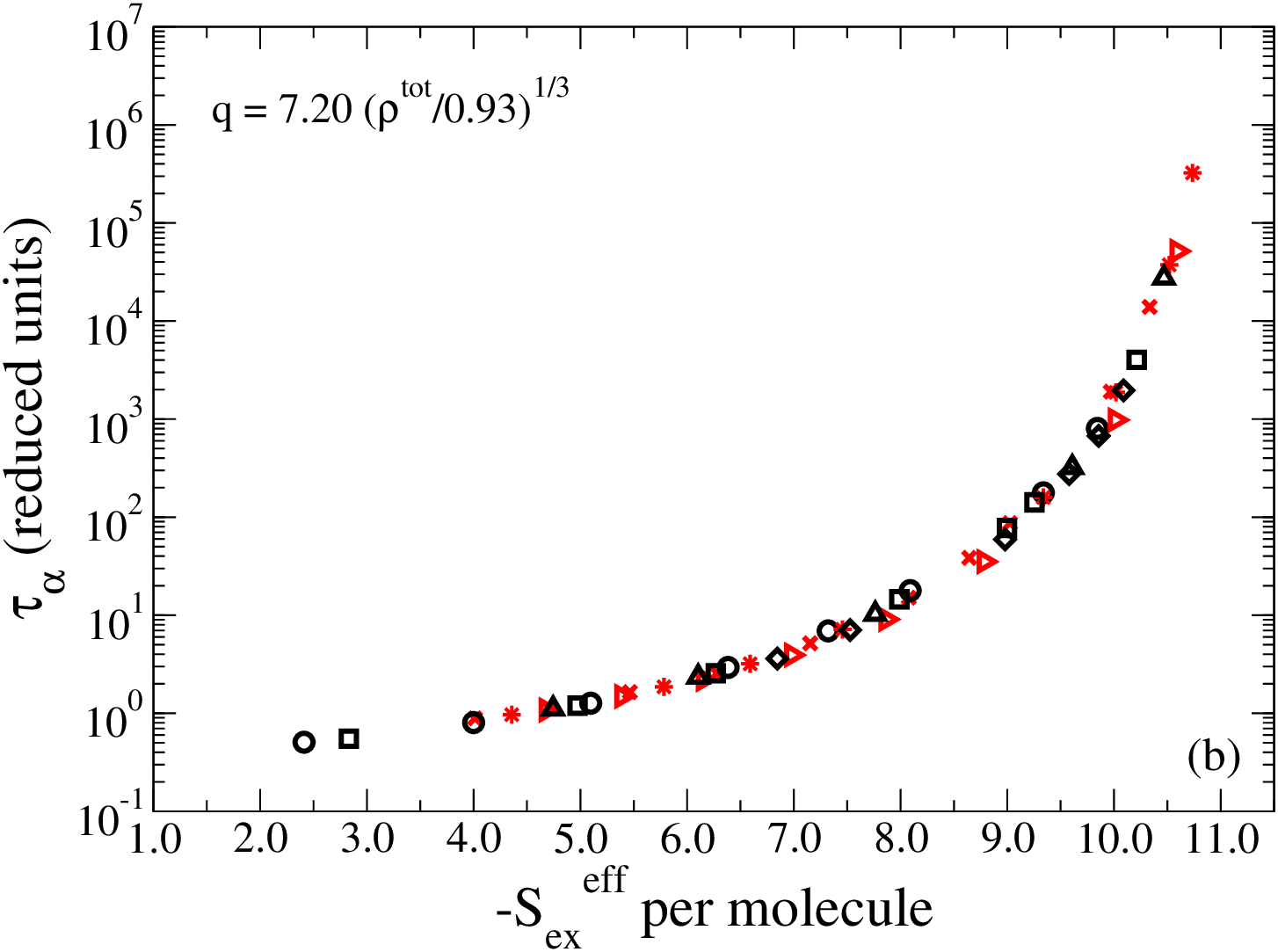}
  \caption{The reduced average structural relaxation time $\tilde{\tau}_{\alpha}$ in bulk (black) and in confinement (red) plotted (a) as a function of total excess entropy, and (b) as a function of effective excess entropy (see the Supplemental Material for definitions). The bulk simulations have $\rho$ = 0.77 and 0.14 $\leq$ T $\leq$ 12.5; $\rho$ = 0.85 and 0.25 $\leq$ T $\leq$ 12.5; $\rho$ = 1.01 and 0.69 $\leq$ T $\leq$ 2.00.}
    \label{excess}
\end{figure}

If the structural relaxation time is plotted against the average density (Figs. \ref{density}(a) and (b)), it is clear that density does not capture the changes that occur going into the highly viscous regime. As an example, comparing at the same average density $\rho^{\rm eff} = 1.1$ (see Fig. \ref{density}(b)), one would predict four decades too fast dynamics for a highly confined system (red crosses) using the bulk behavior (black triangles). The inset of Fig. \ref{density}(b) shows the crucial effect that temperature has on the confined dynamics at conditions typical for this study. A super-Arrhenius behavior is observed at the lowest temperatures, consistent with experimental findings\cite{dosseh2006}. Figures \ref{excess} and \ref{density} show that free-volume-type theories\cite{cohen1979} cannot predict the dynamic consequences of confining the fluid, whereas the more microscopic, correlation-based measure $S_{\rm ex}$ can. 
\newline \newline
\begin{figure}[H]
  \centering
  \includegraphics[width=80mm]{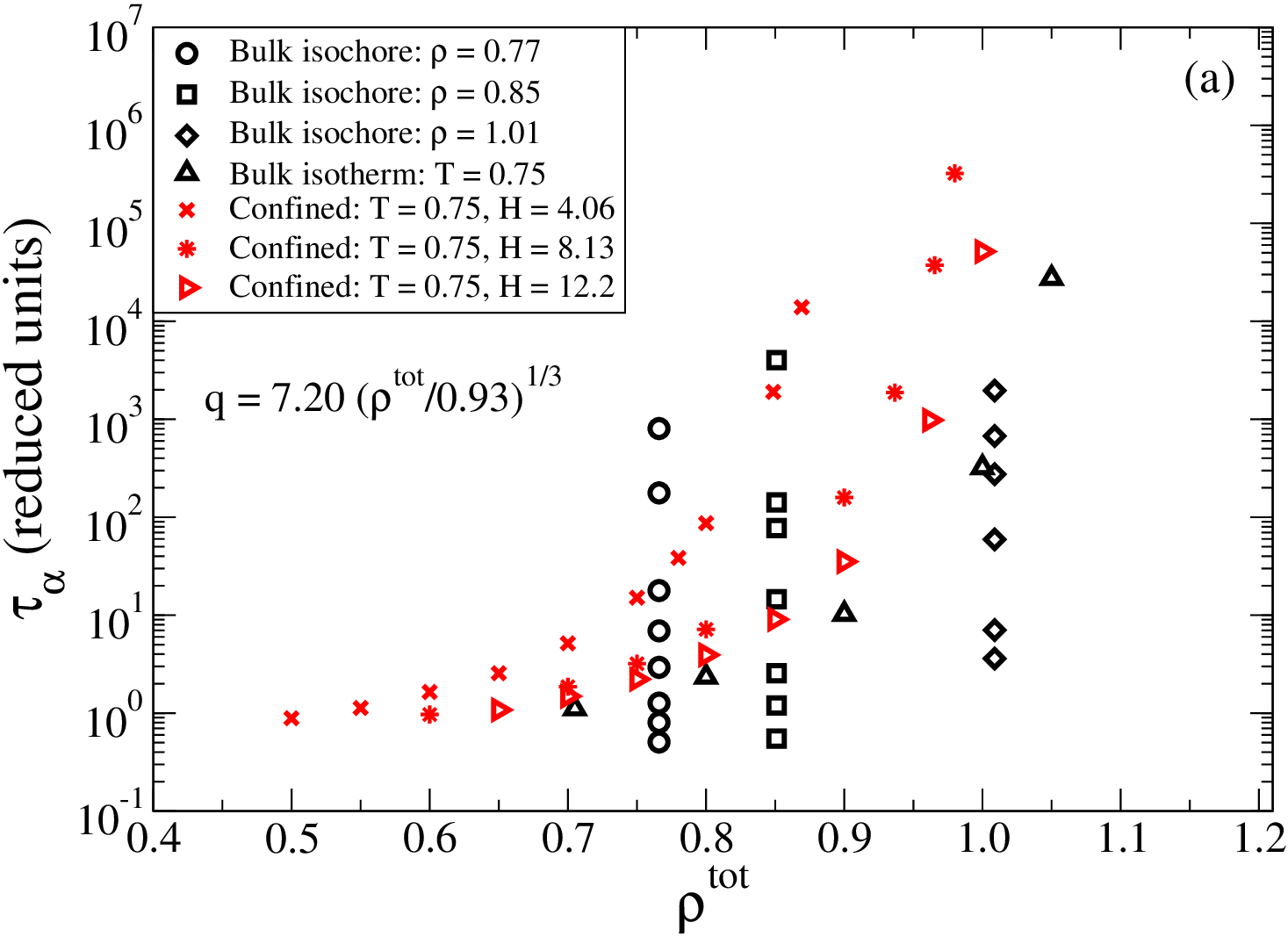}
\end{figure}
\begin{figure}[H]
  \centering
  \includegraphics[width=80mm]{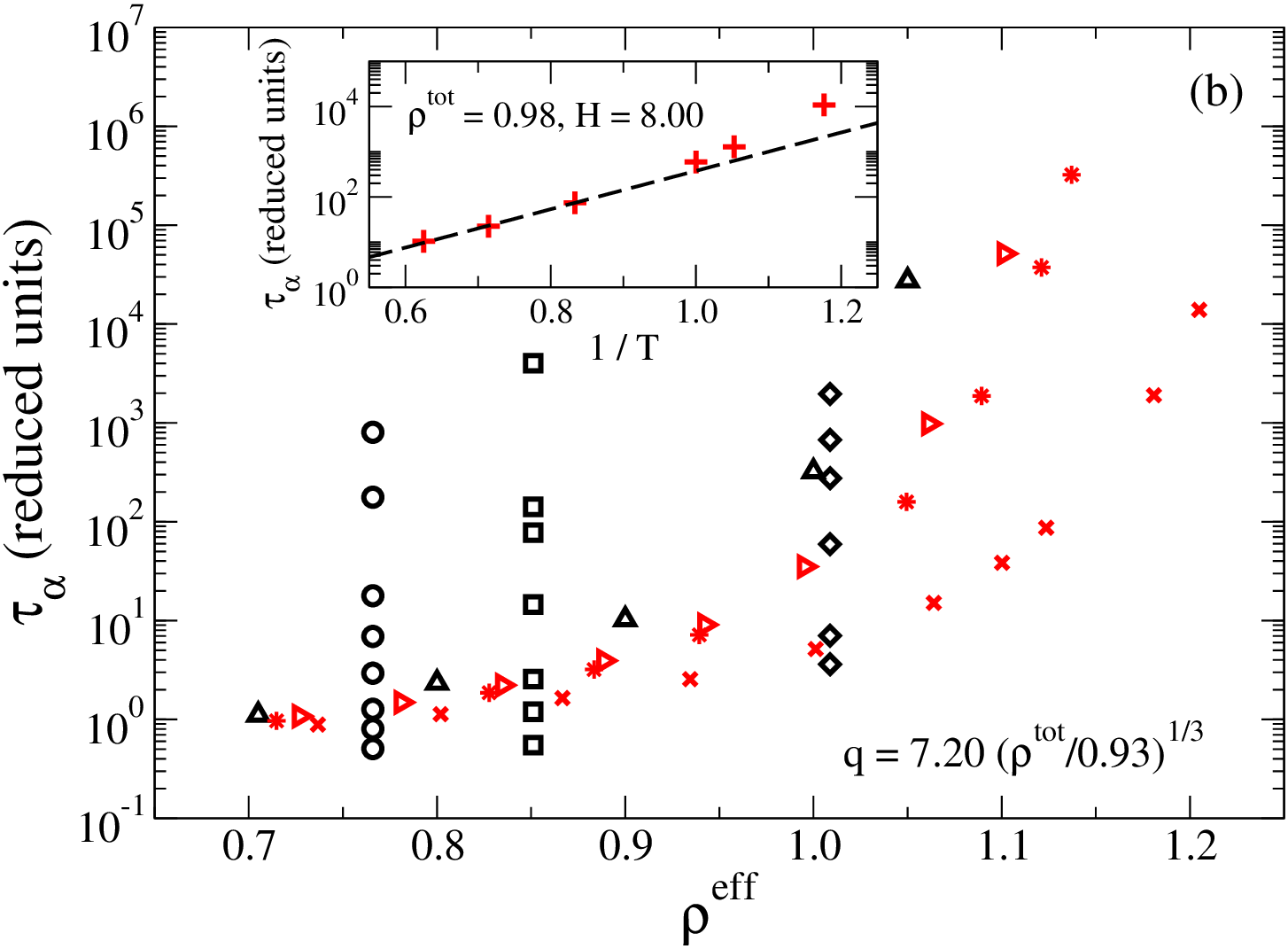}
  \caption{The reduced average structural relaxation time $\tilde{\tau}_{\alpha}$ in confinement (red) and bulk (black) plotted (a) as a function of average density $\rho^{\rm tot} = n/(HA)$, and (b) as a function of effective average slit-pore density $\rho^{\rm eff} = n/(H^{\rm eff}A)$ (see the Supplemental Material), where $n$ is the number of molecules. The inset shows the crucial effect that temperature has on the confined dynamics at conditions typical for this study. The bulk simulations have $\rho$ = 0.77 and 0.14 $\leq$ T $\leq$ 12.5; $\rho$ = 0.85 and 0.25 $\leq$ T $\leq$ 12.5; $\rho$ = 1.01 and 0.69 $\leq$ T $\leq$ 2.00.}
  \label{density}
\end{figure}

\begin{figure}[H]
  \centering
  \includegraphics[width=80mm]{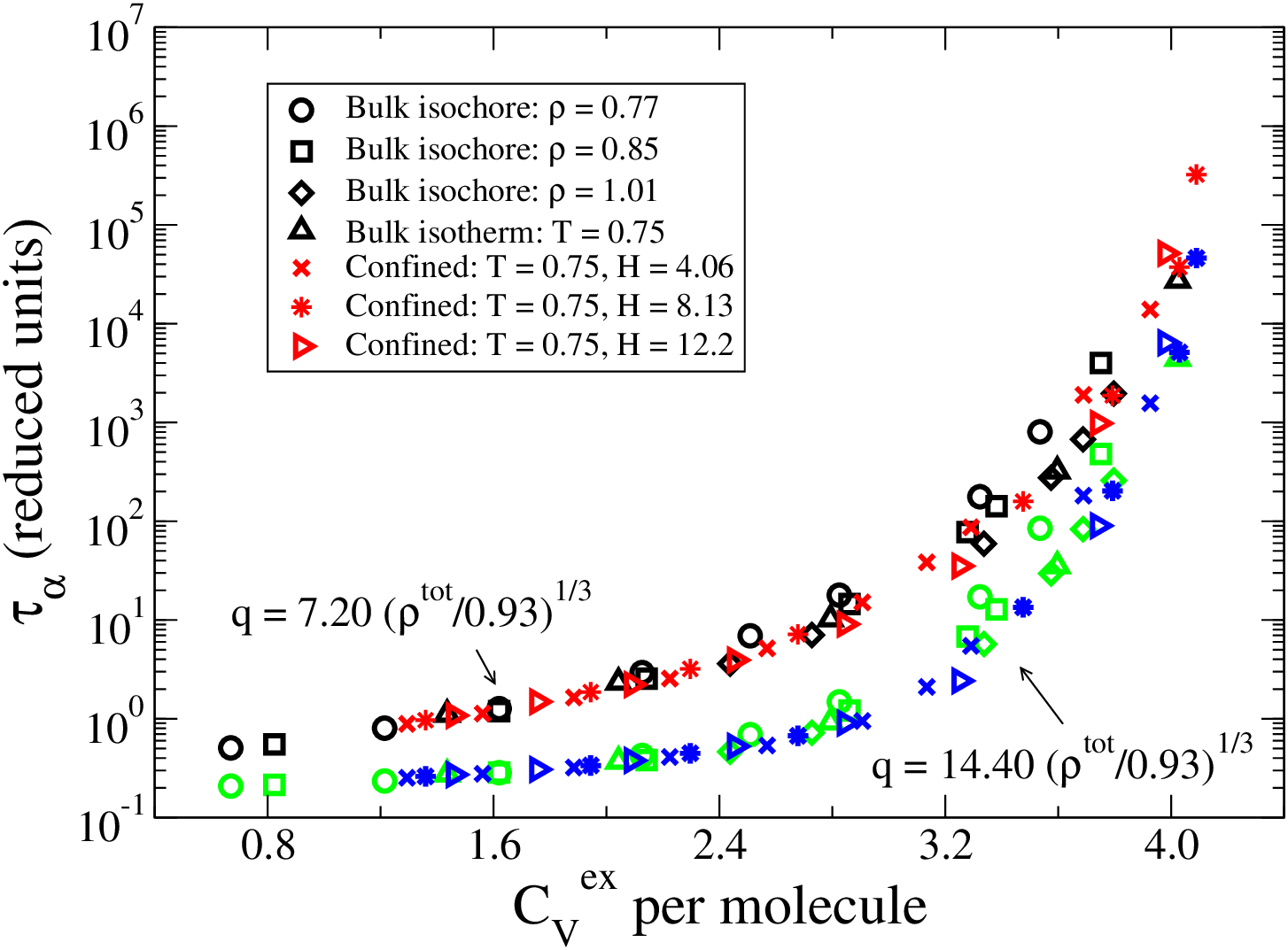}
  \caption{The reduced average structural relaxation time for the same q vector as studied previously, $\tilde{\tau}_{\alpha}$, in bulk (black) and in confinement (red) plotted as a function of the excess isochoric heat capacity $C_{V}^{\rm ex}$ per molecule. To a good approximation the heat capacity, like the entropy, is able to rationalize the dynamical changes induced by confinement. The bulk simulations have $\rho$ = 0.77 and 0.14 $\leq$ T $\leq$ 12.5; $\rho$ = 0.85 and 0.25 $\leq$ T $\leq$ 12.5; $\rho$ = 1.01 and 0.69 $\leq$ T $\leq$ 2.00. To show that the data collapse is not specific for one wavevector, the figure shows bulk (green) and confinement (blue) data points for the double wave vector.}
  \label{heat}
\end{figure}

We probed also a quantity that is much easier to calculate in simulations than $S_{\rm ex}$, namely the excess isochoric heat capacity given by  $C_{V}^{\rm ex} = \langle (\Delta U)^{2}\rangle/k_{B}T^{2}$ ($U$ is the potential energy, $k_B$ Boltzmann's constant, and $T$ the temperature). Figure \ref{heat} shows the structural relaxation time plotted as a function of $C_{V}^{\rm ex}$. This quantity captures the dynamics of confinement over the full time span of six decades, although the collapse is not as good as for the excess entropy. Notably, the relaxation times of bulk and confined systems depend {\it in the same way} on $C_{V}^{\rm ex}$, just as is the case for $S_{\rm ex}$. 
\newline \newline

According to Rosenfeld\cite{rosenfeld1}, the relaxation time is controlled by the excess entropy because a low excess entropy implies that many states are effectively rendered inaccessible, thereby increasing the relaxation time. But why does $C_{V}^{\rm ex}$ also predict the dynamics, why does density not work, and how general may one expect these findings to be? A possible explanation refers to the existence of isomorphs\cite{paper4} in systems that display strong correlations between the equilibrium fluctuations of the potential energy $U$ and the virial $W$ in the \textit{NVT} ensemble\cite{paper1} (``Roskilde simple'' systems\cite{prx}). Recall that the instantaneous energy and  pressure are each sum of a trivial kinetic part and a configurational part. The latter are $U$ and $W$, respectively. At any given state point the Pearson correlation coefficient $R$ for the \textit{NVT} thermal equilibrium fluctuations of $U$ and $W$ measures the strength of the correlations. Only inverse power-law fluids are perfectly correlating ($R=1$), but many models\cite{paper1}, e.g., the Lennard-Jones liquid, and some experimental liquids\cite{gammagamma} have been shown to belong to the class of Roskilde-simple liquids defined by requiring $ R \geq 0.90$\cite{paper1}. This class appears to include most or all van der Waals and metallic liquids, but exclude most or all covalently, hydrogen-bonding, or strongly ionic or dipolar liquids\cite{paper1}. 

Roskilde-simple liquids are characterized by having good isomorphs\cite{paper4}. An isomorph is a curve in the thermodynamic phase diagram along which structure and dynamics are invariant in reduced units; the excess entropy and the excess isochoric heat capacity are also invariant (but not the density). Since the reduced relaxation time is an isomorph invariant, both the excess entropy and the excess isochoric heat capacity predict the dynamics of Roskilde-simple liquids, whereas density does not. 

\begin{figure}[H]
  \centering
  \includegraphics[width=80mm]{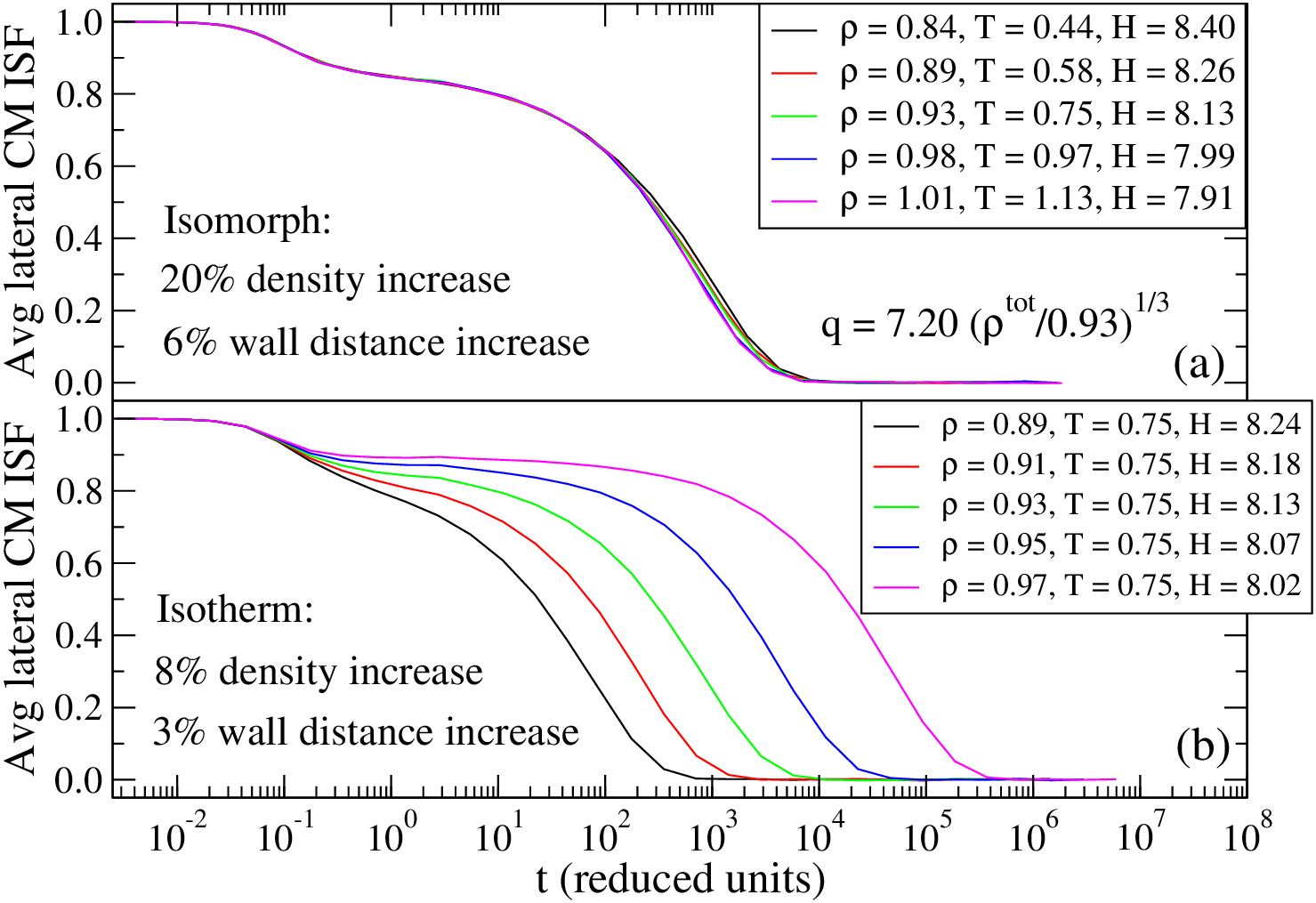}
  \caption{The average molecular center-of-mass incoherent intermediate scattering function as a function of reduced time for various state points of the asymmetric dumbbell model in confinement (a) along a confined isomorph, and (b) along a confined isotherm. The dynamics is to a good approximation invariant along the isomorph, but not along the isotherm even though it involves less than half the density variation.}
  \label{iso}
\end{figure}

In the bulk liquid phase the asymmetric dumbbell model is Roskilde simple\cite{moleculesisomorphs}. To apply isomorph reasoning to a confined system, however, one needs to show that isomorphs exist also for the nanoscale-confined liquid, which has entirely different physics. We document this in Fig. \ref{iso}, where the molecular center-of-mass incoherent intermediate scattering function is shown along an isomorph and an isotherm. The dynamics is to a good approximation invariant along the isomorph, whereas along the isotherm shows a substantial variation for less than half the density variation. We observed a similar behavior when probing the dynamics parallel to the walls at a fixed distance in reduced units from the wall (results not shown). Interestingly, the nanodynamics is isomorph invariant even though it is known to be spatially heterogeneous; this is because the entire spatial relaxation-time distribution in reduced units is predicted to be isomorph invariant.

The Supplemental Material gives details on the definition of isomorphs in confinement and how they are generated in simulation. Briefly, the idea is the following. $H$ is the distance between the two points where the wall potentials diverge, and $A$ the interfacial area of the slit-pore volume. Consider two state points ($H_{1}$, $A_{1}$, $T_{1}$) and ($H_{2}$, $A_{2}$, $T_{2}$) in the phase diagram of a confined liquid for which the state variables are related via ${H^{2}_{1}}/{A_{1}} = {H^{2}_{2}}/{A_{2}}$, implying that a homogenous scaling of space maps slit pore $1$ onto slit pore $2$. These state points are {\it isomorphic} if the following holds: Two microconfigurations, one of each state point, have proportional Boltzmann statistical probabilities whenever they for all molecules $i$ have identical reduced coordinates, i.e., $\rho_{A_{1}}^{1/2}\,x^{(1)}_{CM,i} = \rho_{A_{2}}^{1/2}\,x^{(2)}_{CM,i}$, $\rho_{A_{1}}^{1/2}\,y^{(1)}_{CM,i} = \rho_{A_{2}}^{1/2}\,y^{(2)}_{CM,i}$, $\rho_{H_{1}}\,z^{(1)}_{CM,i} = \rho_{H_{2}}\,z^{(2)}_{CM,i}$ (in which $\rho_{H} \equiv n/H$, $\rho_{A} \equiv n/A$, and $n$ is the number of molecules), as well as identical Eulerian angles. In particular, isomorphic state points are identical in their packing arrangments. If $\textbf{R}$ is the collective configuration space coordinate this means that $\exp(-U(\textbf{R}^{(1)})/k_{B}T_{1}) = C_{12}\exp(-U(\textbf{R}^{(2)})/k_{B}T_{2})$ where $C_{12}$ depends only on the two thermodynamic state points, not on the microconfigurations. Taking the logarithm of this and rearranging, we get

\begin{equation} \label{directiso}
  U(\textbf{R}^{(2)}) = \frac{T_{2}}{T_{1}} U(\textbf{R}^{(1)}) + k_{B}T_{2}\ln C_{12}.
\end{equation}
Isomorphs are generated using this ``direct isomorph check''\cite{paper4} relation, where the walls of slit-pore follow the overall scaling in total density. 
\newline \newline

A very recently empirically established property of Roskilde-simple model liquids is that they obey Rosenfeld-Tarazona (RT) scaling ($C_V^{\rm ex}\propto T^{-2/5}$) significantly better than liquids in general\cite{RTingebrigtsen}. From this one can understand why $C_V^{\rm ex}$ and $S_{\rm ex}$ both collapse the bulk data: Integration of $C_V^{\rm ex}=(\partial S_{\rm ex}/\partial\ln T)_\rho \propto T^{-2/5}$ leads to $-S_{\rm ex}=(5/2)C_V^{\rm ex} + K(\rho)$. Isomorph invariance of $S_{\rm ex}$ and $C_V^{\rm ex}$ implies $K(\rho)=0$, i.e., $-S_{\rm ex}=(5/2)C_V^{\rm ex}$. This is consistent with Figs. 2 and 4, but these figures tell us more, namely that both entropy and specific heat control the relaxation time of the bulk and the confined system {\it in the same way}.  

Recently, Watanabe \textit{et al.}\cite{watanabe2011} showed that the dynamics of a confined fluid system as a function of the distance to the walls can be described to a good approximation using the magnitude of the medium-range crystalline order\cite{shintani2006,kawasaki2007,leocmach2012}. A relation between the two-body excess entropy and the size of these regions has also been reported\cite{kawasaki2007}. The two-body excess entropy is an isomorph invariant\cite{paper4}, so the results of Watanabe \textit{et al.} confirm the existence of isomorphs in confinement.

Theories for confined liquids\cite{krakoviack2005,biroli2006,lang2010,lang2012} must be consistent with the existence of isomorphs for Roskilde-simple fluids, a requirement which may be used as a ``filter'' when developing new approaches\cite{paper4}: Any theory for the reduced relaxation time -- an isomorph invariant -- must express this as a function of another isomorph invariant. Isomorphs are only relevant for fluids that are Roskilde simple\cite{prx}, however. One should not expect the dynamic/thermodynamic correlations to hold for strongly self-associating or network-forming liquids like water, which are not Roskilde simple, even in the bulk. Similarly, such correlations will hardly hold for certain idealized models, e.g., infinitely thin needles or crosses with ideal-gas-like static correlations\cite{ketel2005}, whose slow relaxations at high density are due to topological constraints which, while not reflected in structure, hinder localized dynamic rearrangements. Finally, one expects such correlations also to break down if length and energy scales of the fluid-wall interaction are substantially different from the fluid-fluid interaction or if the confining pores are very narrow. Evidence for the latter can be seen in Fig. 2 of the Supplemental Material.

To summarize, the excess entropy has for a long time been used to describe the relaxation time of liquids\cite{adam1965,rosenfeld1} and more recently shown to work also for confined systems\cite{hsconfined}. We demonstrated above a new controlling variable, the excess isochoric heat capacity, which is expected to apply for the fairly large class of liquids with strong correlations between virial and potential energy fluctuations in the \textit{NVT} ensemble. We welcome new experimental as well as additional simulation studies of a wide spectrum of confined liquids to probe for the existence of isomorphs for confined liquids and, moreover, test the possible generality, beyond Roskilde-simple liquids, of this intriguing relation between static and dynamic properties.

The center for viscous liquid dynamics ``Glass and Time'' is sponsored by the Danish National Research Foundation (DNRF61). Thomas Truskett acknowledges support of the Welch Foundation (F-1696) and the National Science Foundation (CBET-1065357). Jeffrey Errington acknowledges support of the National Science Foundation (CHE-1012356). We also acknowledge the Texas Advanced Computing Center (TACC) at The University of Texas at Austin and the Center for Computational Research at the University at Buffalo for providing HPC resources that have contributed to the research results reported within this paper. Trond S. Ingebrigtsen acknowledges useful discussions of confined-liquid behavior with S{\o}ren Toxvaerd.
\newline \newline \newline
{\Large \textbf{Supplemental Material}}
\newline \newline 
\textbf{Methods}
\newline \newline
\textit{Simulation details} \newline \newline
All simulations were performed (except when calculating the excess entropy)  using a molecular dynamics code optimized for \emph{NVIDIA} graphics cards, which is available as open source code at http://rumd.org. More specifically, we used \textit{NVT} molecular dynamics with $n = 500$ molecules, and generated the \textit{NVT} ensemble via the Nos\'{e}-Hoover method\cite{nose}. 
Only insignificant size-effects were observed by increasing the system size to twice the number of molecules.
The time step used is $\Delta t = 0.0025$ in the unit system defined below. 

The asymmetric dumbbell model\cite{moleculeshidden} consists of a large ($A$) and a small ($B$) Lennard-Jones (LJ) atom, rigidly bonded  with bond distance of $r_{AB} = 0.29/0.4963$ (here and 
henceforth units are given in LJ units referring to the $A$ atom such that $\sigma_{AA}$ = 1, $\epsilon_{AA}$ = 1, and $m_{A}$ = 1). The asymmetric dumbbell 
model has $\sigma_{BB}=0.3910/0.4963$, $\epsilon_{BB}=0.66944/5.726$, and $m_{B}=15.035/77.106$. The $AB$ interaction between different 
molecules is determined by the Lorentz-Berthelot mixing rule.

\begin{figure}[H]
  \centering
  \includegraphics[width=90mm]{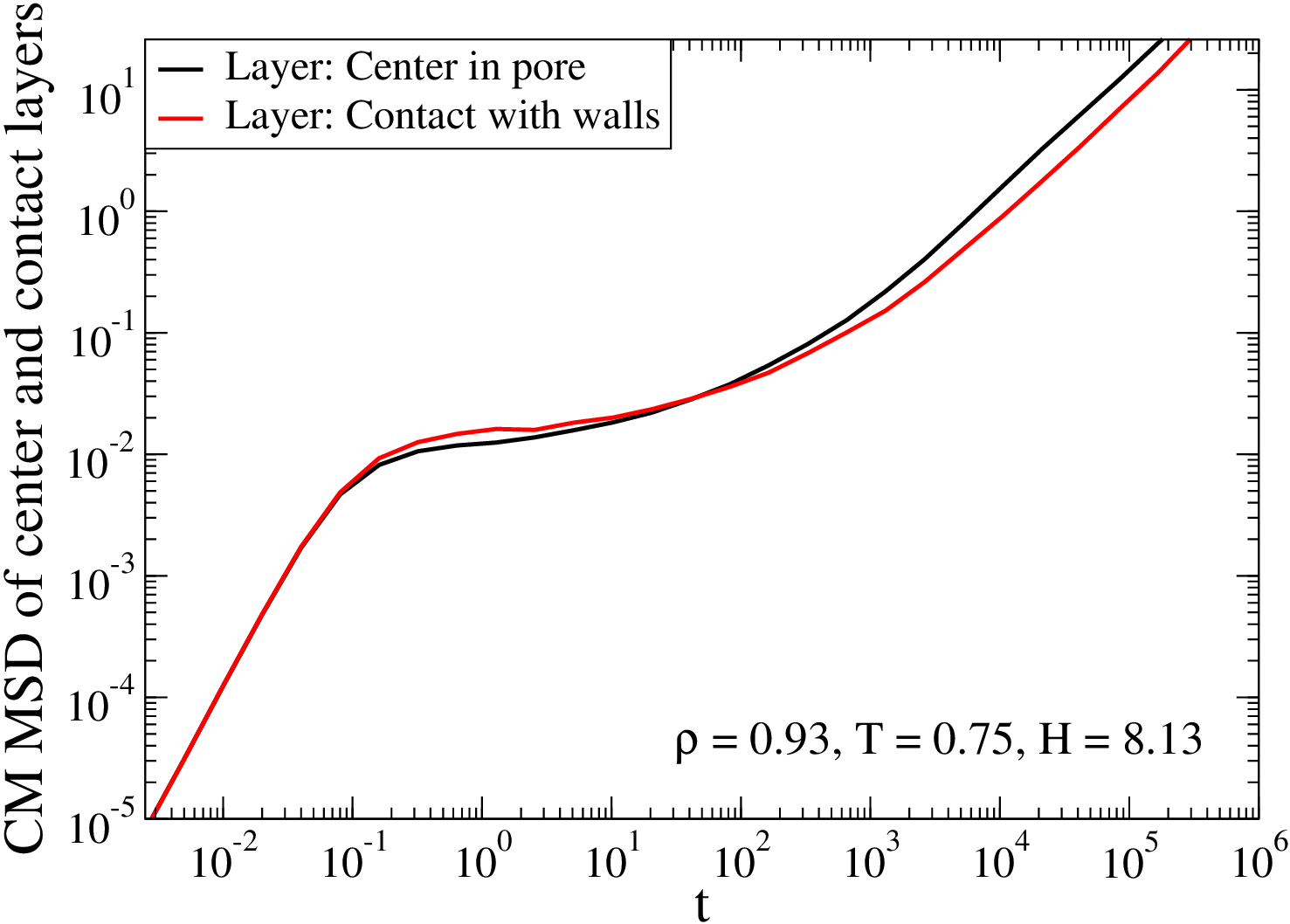}
  \caption{The heterogeneous dynamics of the slit-pore confinement quantified via the lateral center-of-mass (CM) mean-square displacement.}
  \label{hetro}
\end{figure}

The walls are modelled via a smooth potential adding the contribution $U_{WALL} = \sum_{i}\big(u_{9,3}(z_{upper}-z_{i}) + u_{9,3}(z_{i}-z_{lower})\big)$ to the (total) potential energy. 
The external wall potential is given by

\begin{equation}
  u_{9,3}(z) = \frac{4\pi\epsilon_{iw}\rho_{w}\sigma^{3}_{iw}}{3}\Big[\frac{1}{15}\Big(\frac{\sigma_{iw}}{z}\Big)^{9} - \frac{1}{2}\Big(\frac{\sigma_{iw}}{z}\Big)^{3}\Big].
\end{equation}
\newline
Here, $z$ is the distance between the divergence of the potential and the atom in question. $\sigma_{iw}$ and $\epsilon_{iw}$ are parameters similar to the LJ potential, and $\rho_{w} = 0.9316$ defines the density of the confining solid. We set $\sigma_{Aw}$ = 1, $\epsilon_{Aw}$ = 1, $\sigma_{Bw}$ = $(1 + \sigma_{BB})/2$, $\epsilon_{Bw}$ = $\sqrt{ 1 \cdot \epsilon_{BB}}$. The dynamics near the walls and in the center of the slit-pore is shown in Fig. \ref{hetro}.
\newline \newline
\textit{Defining density}
\newline \newline
We define $H$ as the distance between the two points where the wall potentials diverge, i.e.; $H = z_{upper} - z_{lower}$. The total density is then defined as $\rho^{\rm tot} = n /(HA)$, where $n$ is the number of molecules, and $A$ the interfacial area of the slit-pore simulation box volume. The effective slit-pore density $\rho^{\rm eff} = n / (H^{\rm eff}A)$ is estimated following the method outlined in Ref. \onlinecite{dumbbellconfinedroughwalls} for calculating $H^{\rm eff}$. The only difference between this approach and ours is that the absolute minimum of the effective wall-fluid potential is used as a shift. This minimum is present over the full range of state points investigated here. Only insignificant differences between these choices were observed. We have also used other definitions of the average density, and the conclusion of Fig. 3 (main text) remains unchanged.
\newline \newline 
\textit{Excess entropy calculations} \newline \newline
The excess entropy at a given state point has been calculated from the thermodynamic relation $S_{\rm ex} = (U - F_{\rm ex})/ k_{B}T$ (in which $F_{\rm ex}$ is the Helmholtz free energy) using an approach similar to that outlined in Ref. \onlinecite{dumbbellconfinedroughwalls}. We first employ grand canonical transition matrix Monte Carlo simulation to obtain the density dependence of the absolute Helmholtz free energy of the fluid at relatively high temperature. We then use expanded ensemble Monte Carlo simulation to follow the variation in the Helmholtz free energy with temperature, density, and/or pore width. For example, when moving along an isochore we create a series of subensembles with variable $T$ and fixed $n$, $A$, and $H$. The potential energy $U$ is obtained from an ensemble average within the relevant subensemble.  The excess entropy is defined with respect to an uncorrelated ideal gas of either density $\rho^{\rm tot}$ or $\rho^{\rm eff}$ (see density section).  A system of volume $AH$ = 1000 is used to complete grand canonical simulations. All expanded ensemble simulations employ $n$ = 1000 molecules.
\newline \newline
\textit{Isomorph definition and generation in confinement}\newline \newline
We define two state points ($H_{1}$, $A_{1}$, $T_{1}$) and ($H_{2}$, $A_{2}$, $T_{2}$) in the phase diagram of a liquid confined to a slit-pore, where the state variables are related via

\begin{equation}\label{newrel}
  \frac{H^{2}_{1}}{A_{1}} = \frac{H^{2}_{2}}{A_{2}},
\end{equation}
to be isomorphic, if the following holds: Whenever two configurations of state points ($1$) and ($2$) for all molecules $i$ have identical reduced center-of-mass coordinates 
($\rho_{H} \equiv n/H$, $\rho_{A} \equiv n/A$, and $n$ is the number of molecules)
\begin{align}
  \rho_{A_{1}}^{1/2}\,x^{(1)}_{CM,i} & = \rho_{A_{2}}^{1/2}\,x^{(2)}_{CM,i},\label{test1} \\ 
  \ \rho_{A_{1}}^{1/2}\,y^{(1)}_{CM,i} & = \rho_{A_{2}}^{1/2}\,y^{(2)}_{CM,i}, \\ 
  \ \rho_{H_{1}}\,z^{(1)}_{CM,i} & = \rho_{H_{2}}\,z^{(2)}_{CM,i} \label{test3},
\end{align}
and identical Eulerian angles

\begin{align}
  \phi_{i}^{(1)} = \phi_{i}^{(2)}, \ \theta_{i}^{(1)} = \theta_{i}^{(2)}, \ \chi_{i}^{(1)} = \chi_{i}^{(2)}, \label{moleculeiso2}
\end{align}
these two configurations have proportional Boltzmann factors, i.e., [where $\textbf{R}$ $\equiv$ ($\textbf{r}_{CM,1}$, $\phi_{1}$, $\theta_{1}$, $\chi_{1}$, ..., 
$\textbf{r}_{CM,N}$, $\phi_{N}$, $\theta_{N}$, $\chi_{N}$)]

\begin{equation} \label{defisoslit}
  e^{-U(\textbf{R}^{(1)})/k_{B}T_{1}} = C_{12}e^{-U(\textbf{R}^{(2)})/k_{B}T_{2}}.
\end{equation}
Here $C_{12}$ is a constant that depends only on the two thermodynamic state points, not on the microconfigurations. Taking the logarithm of Eq. (\ref{defisoslit}), and rearranging, we get

\begin{equation} \label{directiso1}
  U(\textbf{R}^{(2)}) = \frac{T_{2}}{T_{1}} U(\textbf{R}^{(1)}) + k_{B}T_{2}\ln C_{12}.
\end{equation}
Equation (\ref{directiso1}) is called the ''direct isomorph check''\cite{paper4}. The isomorph is generated using this relation, where the walls of slit-pore follow the overall scaling in \emph{total} density. This works as follows: For each microconfiguration we change $H$ and $A$ according to Eq. (\ref{newrel}) corresponding to a density change of 1\%, scale the microconfiguration accordingly, and calculate the new potential energy. Plotted against the potential energy before scaling, the linear regression slope provides the ratio $T_{2}/T_{1}$; thus the temperature of the isomorphic state point $T_{2}$ is calculated simply by multiplying the slope by $T_{1}$. This procedure is repeated for each state point along the isomorph until a curve in the phase diagram is generated. The starting state point for the generation of the isomorph is $\rho$ = 0.93, $T$ = 0.75, $H$ = 8.13, i.e., the green curve of Fig. 5(a) (main text).

\begin{figure}[H]
  \centering
  \includegraphics[width=90mm]{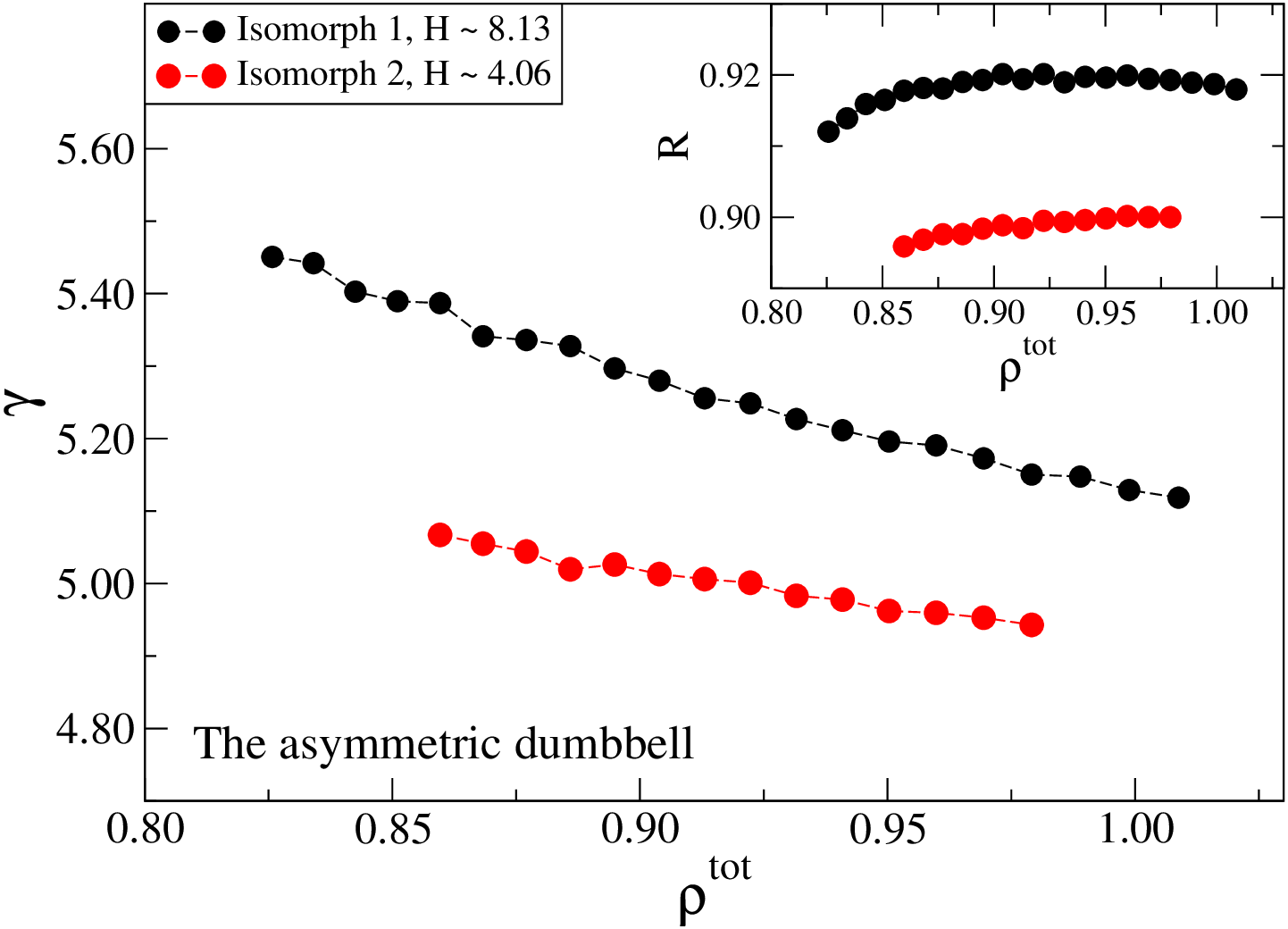}
  \caption{$\gamma = \langle \Delta W \Delta U \rangle / \langle (\Delta U)^{2} \rangle$ as a function of $\rho^{tot}$
    along two isomorphs of the asymmetric dumbbell
    model in a slit-pore (black and red data points). The inset shows  the correlation
    coefficient $R = \langle \Delta W \Delta U \rangle / \sqrt{\langle(\Delta W)^{2}\rangle}\sqrt{\langle (\Delta U)^{2}} \rangle$.}
  \label{risit}
\end{figure}

Figure \ref{risit} shows the density-scaling exponent\cite{thermoscl} $\gamma = \langle \Delta W \Delta U \rangle / \langle (\Delta U)^{2} \rangle$ and the 
correlation coefficient $R = \langle \Delta W \Delta U \rangle / \sqrt{\langle(\Delta W)^{2}\rangle}\sqrt{\langle (\Delta U)^{2}} \rangle$ as 
function of $\rho^{tot}$ for two different isomorphs generated via the method described above. One notes that for $H \approx$ 4.06, the correlation 
coefficient $R$ is significantly less than for $H \approx 8.13$ (see inset) and thus this isomorph is  more approximative. The correlation coefficient is, however, still greater than $0.90$ for most of the investigated state points.

\textit{Analytical methods}
\newline\newline
Except for Fig. 1 (main text), times are measured in dimensionless/reduced units (denoted by a tilde) defined from 
thermodynamic quantities\cite{paper4} via $\tilde{t} = t(\rho^{\rm tot})^{1/3}\sqrt{k_{B}T/\langle m \rangle}$. The reduced molecular 
structural relaxation times $\tilde{\tau}_{\alpha}$ have been extracted from the molecular center-of-mass incoherent intermediate scattering function for a wave vector parallel to the walls. The modulus $q$ of this wave vector is given in the figures. The relaxation times are extracted when $F_{sCM}(\tilde{\tau}_{\alpha}) = 0.2$.

The fixed axis of Fig. 1 (main text) is defined by taking the $z$-axis normal to the walls of slit-pore, and $cos \,\theta$ is calculated from the (ensemble averaged) 
dot-product of a unit-vector in the $z$-direction and a unit-vector connecting the $B$ and $A$ atoms of a molecule. The grid-spacing used is $\Delta z = 0.01$.
\newline \newline

%\bibliography{../../ph_d_thesis/mybib}

\begin{thebibliography}{62}
\expandafter\ifx\csname natexlab\endcsname\relax\def\natexlab#1{#1}\fi
\expandafter\ifx\csname bibnamefont\endcsname\relax
  \def\bibnamefont#1{#1}\fi
\expandafter\ifx\csname bibfnamefont\endcsname\relax
  \def\bibfnamefont#1{#1}\fi
\expandafter\ifx\csname citenamefont\endcsname\relax
  \def\citenamefont#1{#1}\fi
\expandafter\ifx\csname url\endcsname\relax
  \def\url#1{\texttt{#1}}\fi
\expandafter\ifx\csname urlprefix\endcsname\relax\def\urlprefix{URL }\fi
\providecommand{\bibinfo}[2]{#2}
\providecommand{\eprint}[2][]{\url{#2}}

\bibitem[{\citenamefont{Bhushan et~al.}(1995)\citenamefont{Bhushan,
  Israelachvili, and Landman}}]{bhushan1995}
\bibinfo{author}{\bibfnamefont{B.}~\bibnamefont{Bhushan}},
  \bibinfo{author}{\bibfnamefont{J.~N.} \bibnamefont{Israelachvili}},
  \bibnamefont{and} \bibinfo{author}{\bibfnamefont{U.}~\bibnamefont{Landman}},
  \bibinfo{journal}{Nature} \textbf{\bibinfo{volume}{374}},
  \bibinfo{pages}{607} (\bibinfo{year}{1995}).

\bibitem[{\citenamefont{Whitesides}(2006)}]{whitesides2006}
\bibinfo{author}{\bibfnamefont{G.~M.} \bibnamefont{Whitesides}},
  \bibinfo{journal}{Nature} \textbf{\bibinfo{volume}{442}},
  \bibinfo{pages}{368} (\bibinfo{year}{2006}).

\bibitem[{\citenamefont{Drake and Klafter}(1990)}]{drake1990}
\bibinfo{author}{\bibfnamefont{J.~M.} \bibnamefont{Drake}} \bibnamefont{and}
  \bibinfo{author}{\bibfnamefont{J.}~\bibnamefont{Klafter}},
  \bibinfo{journal}{Phys. Today} \textbf{\bibinfo{volume}{43}},
  \bibinfo{pages}{46} (\bibinfo{year}{1990}).

\bibitem[{\citenamefont{Granick}(1991)}]{granick1991}
\bibinfo{author}{\bibfnamefont{S.}~\bibnamefont{Granick}},
  \bibinfo{journal}{Science} \textbf{\bibinfo{volume}{253}},
  \bibinfo{pages}{1374} (\bibinfo{year}{1991}).

\bibitem[{\citenamefont{Morineau et~al.}(2002)\citenamefont{Morineau, Xia, and
  Alba-Simionesco}}]{morineau2002}
\bibinfo{author}{\bibfnamefont{D.}~\bibnamefont{Morineau}},
  \bibinfo{author}{\bibfnamefont{Y.}~\bibnamefont{Xia}}, \bibnamefont{and}
  \bibinfo{author}{\bibfnamefont{C.}~\bibnamefont{Alba-Simionesco}},
  \bibinfo{journal}{J. Chem. Phys.} \textbf{\bibinfo{volume}{117}},
  \bibinfo{pages}{8966} (\bibinfo{year}{2002}).

\bibitem[{\citenamefont{Jackson and McKenna}(1991)}]{jackson1991}
\bibinfo{author}{\bibfnamefont{C.~L.} \bibnamefont{Jackson}} \bibnamefont{and}
  \bibinfo{author}{\bibfnamefont{G.~B.} \bibnamefont{McKenna}},
  \bibinfo{journal}{J. Non-Cryst. Solids} \textbf{\bibinfo{volume}{131}},
  \bibinfo{pages}{221} (\bibinfo{year}{1991}).

\bibitem[{\citenamefont{Teboul and Alba-Simionesco}(2005)}]{teboul2005}
\bibinfo{author}{\bibfnamefont{V.}~\bibnamefont{Teboul}} \bibnamefont{and}
  \bibinfo{author}{\bibfnamefont{C.}~\bibnamefont{Alba-Simionesco}},
  \bibinfo{journal}{Chem. Phys.} \textbf{\bibinfo{volume}{317}},
  \bibinfo{pages}{245} (\bibinfo{year}{2005}).

\bibitem[{\citenamefont{Alcoutlabi and McKenna}(2005)}]{alcoutlabi2005}
\bibinfo{author}{\bibfnamefont{M.}~\bibnamefont{Alcoutlabi}} \bibnamefont{and}
  \bibinfo{author}{\bibfnamefont{G.~B.} \bibnamefont{McKenna}},
  \bibinfo{journal}{J. Phys.: Condens. Matter} \textbf{\bibinfo{volume}{17}},
  \bibinfo{pages}{461} (\bibinfo{year}{2005}).

\bibitem[{\citenamefont{Coasne et~al.}(2011)\citenamefont{Coasne,
  Alba-Simionesco, Audonnet, Dosseh, and Gubbins}}]{coasne2011}
\bibinfo{author}{\bibfnamefont{B.}~\bibnamefont{Coasne}},
  \bibinfo{author}{\bibfnamefont{C.}~\bibnamefont{Alba-Simionesco}},
  \bibinfo{author}{\bibfnamefont{F.}~\bibnamefont{Audonnet}},
  \bibinfo{author}{\bibfnamefont{G.}~\bibnamefont{Dosseh}}, \bibnamefont{and}
  \bibinfo{author}{\bibfnamefont{K.~E.} \bibnamefont{Gubbins}},
  \bibinfo{journal}{Phys. Chem. Chem. Phys.} \textbf{\bibinfo{volume}{13}},
  \bibinfo{pages}{3748} (\bibinfo{year}{2011}).

\bibitem[{\citenamefont{Richert}(2011)}]{richert2011}
\bibinfo{author}{\bibfnamefont{R.}~\bibnamefont{Richert}},
  \bibinfo{journal}{Annu. Rev. Phys. Chem.} \textbf{\bibinfo{volume}{62}},
  \bibinfo{pages}{65} (\bibinfo{year}{2011}).

\bibitem[{\citenamefont{Keddie et~al.}(1994)\citenamefont{Keddie, Jones, and
  Cory}}]{keddie1994}
\bibinfo{author}{\bibfnamefont{J.~L.} \bibnamefont{Keddie}},
  \bibinfo{author}{\bibfnamefont{R.~A.~L.} \bibnamefont{Jones}},
  \bibnamefont{and} \bibinfo{author}{\bibfnamefont{R.~A.} \bibnamefont{Cory}},
  \bibinfo{journal}{Europhys. Lett.} \textbf{\bibinfo{volume}{27}},
  \bibinfo{pages}{59} (\bibinfo{year}{1994}).

\bibitem[{\citenamefont{Serghei et~al.}(2006)\citenamefont{Serghei, Tress, and
  Kremer}}]{sergei2006}
\bibinfo{author}{\bibfnamefont{A.}~\bibnamefont{Serghei}},
  \bibinfo{author}{\bibfnamefont{M.}~\bibnamefont{Tress}}, \bibnamefont{and}
  \bibinfo{author}{\bibfnamefont{F.}~\bibnamefont{Kremer}},
  \bibinfo{journal}{Macromolecules} \textbf{\bibinfo{volume}{39}},
  \bibinfo{pages}{9385} (\bibinfo{year}{2006}).

\bibitem[{\citenamefont{Forrest and Dalnoki-Veress}(2001)}]{forrest2001}
\bibinfo{author}{\bibfnamefont{J.~A.} \bibnamefont{Forrest}} \bibnamefont{and}
  \bibinfo{author}{\bibfnamefont{K.}~\bibnamefont{Dalnoki-Veress}},
  \bibinfo{journal}{Adv. Colloid Interface Sci.} \textbf{\bibinfo{volume}{94}},
  \bibinfo{pages}{167} (\bibinfo{year}{2001}).

\bibitem[{\citenamefont{Ellison and Torkelson}(2003)}]{ellison2003}
\bibinfo{author}{\bibfnamefont{C.~J.} \bibnamefont{Ellison}} \bibnamefont{and}
  \bibinfo{author}{\bibfnamefont{J.~M.} \bibnamefont{Torkelson}},
  \bibinfo{journal}{Nat. Mater.} \textbf{\bibinfo{volume}{2}},
  \bibinfo{pages}{695} (\bibinfo{year}{2003}).

\bibitem[{\citenamefont{Rittigstein et~al.}(2007)\citenamefont{Rittigstein,
  Priestley, Broadbelt, and Torkelson}}]{rittigstein2007}
\bibinfo{author}{\bibfnamefont{P.}~\bibnamefont{Rittigstein}},
  \bibinfo{author}{\bibfnamefont{R.~D.} \bibnamefont{Priestley}},
  \bibinfo{author}{\bibfnamefont{L.~J.} \bibnamefont{Broadbelt}},
  \bibnamefont{and} \bibinfo{author}{\bibfnamefont{J.~M.}
  \bibnamefont{Torkelson}}, \bibinfo{journal}{Nat. Mater.}
  \textbf{\bibinfo{volume}{6}}, \bibinfo{pages}{278} (\bibinfo{year}{2007}).

\bibitem[{\citenamefont{Paeng et~al.}(2012)\citenamefont{Paeng, Richert, and
  Ediger}}]{paeng2012}
\bibinfo{author}{\bibfnamefont{K.}~\bibnamefont{Paeng}},
  \bibinfo{author}{\bibfnamefont{R.}~\bibnamefont{Richert}}, \bibnamefont{and}
  \bibinfo{author}{\bibfnamefont{M.~D.} \bibnamefont{Ediger}},
  \bibinfo{journal}{Soft Matter} \textbf{\bibinfo{volume}{8}},
  \bibinfo{pages}{819} (\bibinfo{year}{2012}).

\bibitem[{\citenamefont{Iacob et~al.}(2012)\citenamefont{Iacob, Sangoro,
  Kipnusu, Valiullin, K{\"a}rger, and Kremer}}]{iacob2012}
\bibinfo{author}{\bibfnamefont{C.}~\bibnamefont{Iacob}},
  \bibinfo{author}{\bibfnamefont{J.~R.} \bibnamefont{Sangoro}},
  \bibinfo{author}{\bibfnamefont{W.~K.} \bibnamefont{Kipnusu}},
  \bibinfo{author}{\bibfnamefont{R.}~\bibnamefont{Valiullin}},
  \bibinfo{author}{\bibfnamefont{J.}~\bibnamefont{K{\"a}rger}},
  \bibnamefont{and} \bibinfo{author}{\bibfnamefont{F.}~\bibnamefont{Kremer}},
  \bibinfo{journal}{Soft Matter} \textbf{\bibinfo{volume}{8}},
  \bibinfo{pages}{289} (\bibinfo{year}{2012}).

\bibitem[{\citenamefont{Ji et~al.}(2009)\citenamefont{Ji, Lefort, Ghoufi, and
  Morineau}}]{ji2009}
\bibinfo{author}{\bibfnamefont{Q.}~\bibnamefont{Ji}},
  \bibinfo{author}{\bibfnamefont{R.}~\bibnamefont{Lefort}},
  \bibinfo{author}{\bibfnamefont{A.}~\bibnamefont{Ghoufi}}, \bibnamefont{and}
  \bibinfo{author}{\bibfnamefont{D.}~\bibnamefont{Morineau}},
  \bibinfo{journal}{Chem. Phys. Lett.} \textbf{\bibinfo{volume}{482}},
  \bibinfo{pages}{234} (\bibinfo{year}{2009}).

\bibitem[{\citenamefont{Nugent et~al.}(2007)\citenamefont{Nugent, Edmond,
  Patel, and Weeks}}]{nugent2007}
\bibinfo{author}{\bibfnamefont{C.~R.} \bibnamefont{Nugent}},
  \bibinfo{author}{\bibfnamefont{K.~V.} \bibnamefont{Edmond}},
  \bibinfo{author}{\bibfnamefont{H.~N.} \bibnamefont{Patel}}, \bibnamefont{and}
  \bibinfo{author}{\bibfnamefont{E.~R.} \bibnamefont{Weeks}},
  \bibinfo{journal}{Phys. Rev. Lett.} \textbf{\bibinfo{volume}{99}},
  \bibinfo{pages}{025702} (\bibinfo{year}{2007}).

\bibitem[{\citenamefont{Eral et~al.}(2009)\citenamefont{Eral, van~den Ende,
  Mugele, and Duits}}]{eral2009}
\bibinfo{author}{\bibfnamefont{H.~B.} \bibnamefont{Eral}},
  \bibinfo{author}{\bibfnamefont{D.}~\bibnamefont{van~den Ende}},
  \bibinfo{author}{\bibfnamefont{F.}~\bibnamefont{Mugele}}, \bibnamefont{and}
  \bibinfo{author}{\bibfnamefont{M.~H.~G.} \bibnamefont{Duits}},
  \bibinfo{journal}{Phys. Rev. E} \textbf{\bibinfo{volume}{80}},
  \bibinfo{pages}{061403} (\bibinfo{year}{2009}).

\bibitem[{\citenamefont{Michailidou et~al.}(2009)\citenamefont{Michailidou,
  Petekidis, Swan, and Brady}}]{michailidou2009}
\bibinfo{author}{\bibfnamefont{V.~N.} \bibnamefont{Michailidou}},
  \bibinfo{author}{\bibfnamefont{G.}~\bibnamefont{Petekidis}},
  \bibinfo{author}{\bibfnamefont{J.~W.} \bibnamefont{Swan}}, \bibnamefont{and}
  \bibinfo{author}{\bibfnamefont{J.~F.} \bibnamefont{Brady}},
  \bibinfo{journal}{Phys. Rev. Lett.} \textbf{\bibinfo{volume}{102}},
  \bibinfo{pages}{068302} (\bibinfo{year}{2009}).

\bibitem[{\citenamefont{Watanabe et~al.}(2011)\citenamefont{Watanabe, Kawasaki,
  and Tanaka}}]{watanabe2011}
\bibinfo{author}{\bibfnamefont{K.}~\bibnamefont{Watanabe}},
  \bibinfo{author}{\bibfnamefont{T.}~\bibnamefont{Kawasaki}}, \bibnamefont{and}
  \bibinfo{author}{\bibfnamefont{H.}~\bibnamefont{Tanaka}},
  \bibinfo{journal}{Nat. Mater.} \textbf{\bibinfo{volume}{10}},
  \bibinfo{pages}{512} (\bibinfo{year}{2011}).

\bibitem[{\citenamefont{Edmond et~al.}(2012)\citenamefont{Edmond, Nugent, and
  Weeks}}]{edmond2012}
\bibinfo{author}{\bibfnamefont{K.~V.} \bibnamefont{Edmond}},
  \bibinfo{author}{\bibfnamefont{C.~R.} \bibnamefont{Nugent}},
  \bibnamefont{and} \bibinfo{author}{\bibfnamefont{E.~R.} \bibnamefont{Weeks}},
  \bibinfo{journal}{Phys. Rev. E} \textbf{\bibinfo{volume}{85}},
  \bibinfo{pages}{041401} (\bibinfo{year}{2012}).

\bibitem[{\citenamefont{Fehr and L{\"o}wen}(1995)}]{fehr1995}
\bibinfo{author}{\bibfnamefont{T.}~\bibnamefont{Fehr}} \bibnamefont{and}
  \bibinfo{author}{\bibfnamefont{H.}~\bibnamefont{L{\"o}wen}},
  \bibinfo{journal}{Phys. Rev. E} \textbf{\bibinfo{volume}{52}},
  \bibinfo{pages}{4016} (\bibinfo{year}{1995}).

\bibitem[{\citenamefont{Torres et~al.}(2000)\citenamefont{Torres, Nealey, and
  de~Pablo}}]{torres2000}
\bibinfo{author}{\bibfnamefont{J.~A.} \bibnamefont{Torres}},
  \bibinfo{author}{\bibfnamefont{P.~F.} \bibnamefont{Nealey}},
  \bibnamefont{and} \bibinfo{author}{\bibfnamefont{J.~J.}
  \bibnamefont{de~Pablo}}, \bibinfo{journal}{Phys. Rev. Lett.}
  \textbf{\bibinfo{volume}{85}}, \bibinfo{pages}{3221} (\bibinfo{year}{2000}).

\bibitem[{\citenamefont{Scheidler et~al.}(2000)\citenamefont{Scheidler, Kob,
  and Binder}}]{scheidler2000}
\bibinfo{author}{\bibfnamefont{P.}~\bibnamefont{Scheidler}},
  \bibinfo{author}{\bibfnamefont{W.}~\bibnamefont{Kob}}, \bibnamefont{and}
  \bibinfo{author}{\bibfnamefont{K.}~\bibnamefont{Binder}},
  \bibinfo{journal}{J. Phys. IV France} \textbf{\bibinfo{volume}{10}},
  \bibinfo{pages}{33} (\bibinfo{year}{2000}).

\bibitem[{\citenamefont{Starr et~al.}(2002)\citenamefont{Starr, Schr{\o}der,
  and Glotzer}}]{starr2002}
\bibinfo{author}{\bibfnamefont{F.~W.} \bibnamefont{Starr}},
  \bibinfo{author}{\bibfnamefont{T.~B.} \bibnamefont{Schr{\o}der}},
  \bibnamefont{and} \bibinfo{author}{\bibfnamefont{S.~C.}
  \bibnamefont{Glotzer}}, \bibinfo{journal}{Macromolecules}
  \textbf{\bibinfo{volume}{35}}, \bibinfo{pages}{4481} (\bibinfo{year}{2002}).

\bibitem[{\citenamefont{Baljon et~al.}(2005)\citenamefont{Baljon, Weert,
  DeGraaff, and Khare}}]{baljon2005}
\bibinfo{author}{\bibfnamefont{A.~R.~C.} \bibnamefont{Baljon}},
  \bibinfo{author}{\bibfnamefont{M.~H. M.~V.} \bibnamefont{Weert}},
  \bibinfo{author}{\bibfnamefont{R.~B.} \bibnamefont{DeGraaff}},
  \bibnamefont{and} \bibinfo{author}{\bibfnamefont{R.}~\bibnamefont{Khare}},
  \bibinfo{journal}{Macromolecules} \textbf{\bibinfo{volume}{38}},
  \bibinfo{pages}{2391} (\bibinfo{year}{2005}).

\bibitem[{\citenamefont{Kurzidim et~al.}(2009)\citenamefont{Kurzidim,
  Coslovich, and Kahl}}]{kurzidim2009}
\bibinfo{author}{\bibfnamefont{J.}~\bibnamefont{Kurzidim}},
  \bibinfo{author}{\bibfnamefont{D.}~\bibnamefont{Coslovich}},
  \bibnamefont{and} \bibinfo{author}{\bibfnamefont{G.}~\bibnamefont{Kahl}},
  \bibinfo{journal}{Phys. Rev. Lett.} \textbf{\bibinfo{volume}{103}},
  \bibinfo{pages}{138303} (\bibinfo{year}{2009}).

\bibitem[{\citenamefont{Starr and Douglas}(2011)}]{starr2011}
\bibinfo{author}{\bibfnamefont{F.~W.} \bibnamefont{Starr}} \bibnamefont{and}
  \bibinfo{author}{\bibfnamefont{J.~F.} \bibnamefont{Douglas}},
  \bibinfo{journal}{Phys. Rev. Lett.} \textbf{\bibinfo{volume}{106}},
  \bibinfo{pages}{115702} (\bibinfo{year}{2011}).

\bibitem[{\citenamefont{Betancourt et~al.}(2013)\citenamefont{Betancourt,
  Douglas, and Starr}}]{betancourt2013}
\bibinfo{author}{\bibfnamefont{B.~A.~P.} \bibnamefont{Betancourt}},
  \bibinfo{author}{\bibfnamefont{J.~F.} \bibnamefont{Douglas}},
  \bibnamefont{and} \bibinfo{author}{\bibfnamefont{F.~W.} \bibnamefont{Starr}},
  \bibinfo{journal}{Soft Matter} \textbf{\bibinfo{volume}{9}},
  \bibinfo{pages}{241} (\bibinfo{year}{2013}).

\bibitem[{\citenamefont{Alba-Simionesco
  et~al.}(2006)\citenamefont{Alba-Simionesco, Coasne, Dosseh, Dudziak, Gubbins,
  Radhakrishnan, and Sliwinska-Bartkowiak}}]{alba2006}
\bibinfo{author}{\bibfnamefont{C.}~\bibnamefont{Alba-Simionesco}},
  \bibinfo{author}{\bibfnamefont{B.}~\bibnamefont{Coasne}},
  \bibinfo{author}{\bibfnamefont{G.}~\bibnamefont{Dosseh}},
  \bibinfo{author}{\bibfnamefont{G.}~\bibnamefont{Dudziak}},
  \bibinfo{author}{\bibfnamefont{K.~E.} \bibnamefont{Gubbins}},
  \bibinfo{author}{\bibfnamefont{R.}~\bibnamefont{Radhakrishnan}},
  \bibnamefont{and}
  \bibinfo{author}{\bibfnamefont{M.}~\bibnamefont{Sliwinska-Bartkowiak}},
  \bibinfo{journal}{J. Phys.: Condens. Matter} \textbf{\bibinfo{volume}{18}},
  \bibinfo{pages}{15} (\bibinfo{year}{2006}).

\bibitem[{\citenamefont{Mittal et~al.}(2006)\citenamefont{Mittal, Errington,
  and Truskett}}]{hsconfined}
\bibinfo{author}{\bibfnamefont{J.}~\bibnamefont{Mittal}},
  \bibinfo{author}{\bibfnamefont{J.~R.} \bibnamefont{Errington}},
  \bibnamefont{and} \bibinfo{author}{\bibfnamefont{T.~M.}
  \bibnamefont{Truskett}}, \bibinfo{journal}{Phys. Rev. Lett.}
  \textbf{\bibinfo{volume}{96}}, \bibinfo{pages}{177804}
  (\bibinfo{year}{2006}).

\bibitem[{\citenamefont{Mittal et~al.}(2007)\citenamefont{Mittal, Errington,
  and Truskett}}]{LJconfinedsmoothwalls}
\bibinfo{author}{\bibfnamefont{J.}~\bibnamefont{Mittal}},
  \bibinfo{author}{\bibfnamefont{J.~R.} \bibnamefont{Errington}},
  \bibnamefont{and} \bibinfo{author}{\bibfnamefont{T.~M.}
  \bibnamefont{Truskett}}, \bibinfo{journal}{J. Phys. Chem. B}
  \textbf{\bibinfo{volume}{111}}, \bibinfo{pages}{10054}
  (\bibinfo{year}{2007}).

\bibitem[{\citenamefont{Goel et~al.}(2009)\citenamefont{Goel, Krekelberg, Pond,
  Mittal, Shen, Errington, and Truskett}}]{goel2009}
\bibinfo{author}{\bibfnamefont{G.}~\bibnamefont{Goel}},
  \bibinfo{author}{\bibfnamefont{W.~P.} \bibnamefont{Krekelberg}},
  \bibinfo{author}{\bibfnamefont{M.~J.} \bibnamefont{Pond}},
  \bibinfo{author}{\bibfnamefont{J.}~\bibnamefont{Mittal}},
  \bibinfo{author}{\bibfnamefont{V.~K.} \bibnamefont{Shen}},
  \bibinfo{author}{\bibfnamefont{J.~R.} \bibnamefont{Errington}},
  \bibnamefont{and} \bibinfo{author}{\bibfnamefont{T.~M.}
  \bibnamefont{Truskett}}, \bibinfo{journal}{J. Stat. Mech. Theor. Exp.}
  \textbf{\bibinfo{volume}{4}}, \bibinfo{pages}{04006} (\bibinfo{year}{2009}).

\bibitem[{\citenamefont{Chopra et~al.}(2010)\citenamefont{Chopra, Truskett, and
  Errington}}]{dumbbellconfinedroughwalls}
\bibinfo{author}{\bibfnamefont{R.}~\bibnamefont{Chopra}},
  \bibinfo{author}{\bibfnamefont{T.~M.} \bibnamefont{Truskett}},
  \bibnamefont{and} \bibinfo{author}{\bibfnamefont{J.~R.}
  \bibnamefont{Errington}}, \bibinfo{journal}{Phys. Rev. E}
  \textbf{\bibinfo{volume}{82}}, \bibinfo{pages}{041201}
  (\bibinfo{year}{2010}).

\bibitem[{\citenamefont{Borah et~al.}(2012)\citenamefont{Borah, Maiti,
  Chakravarty, and Yashonath}}]{borah2012}
\bibinfo{author}{\bibfnamefont{B.~J.} \bibnamefont{Borah}},
  \bibinfo{author}{\bibfnamefont{P.~K.} \bibnamefont{Maiti}},
  \bibinfo{author}{\bibfnamefont{C.}~\bibnamefont{Chakravarty}},
  \bibnamefont{and}
  \bibinfo{author}{\bibfnamefont{S.}~\bibnamefont{Yashonath}},
  \bibinfo{journal}{J. Chem. Phys.} \textbf{\bibinfo{volume}{136}},
  \bibinfo{pages}{174510} (\bibinfo{year}{2012}).

\bibitem[{\citenamefont{Gulati and Hall}(1997)}]{gulati1997}
\bibinfo{author}{\bibfnamefont{H.~S.} \bibnamefont{Gulati}} \bibnamefont{and}
  \bibinfo{author}{\bibfnamefont{C.~K.} \bibnamefont{Hall}},
  \bibinfo{journal}{J. Chem. Phys.} \textbf{\bibinfo{volume}{107}},
  \bibinfo{pages}{3930} (\bibinfo{year}{1997}).

\bibitem[{\citenamefont{Schr{\o}der et~al.}(2009)\citenamefont{Schr{\o}der,
  Pedersen, Bailey, Toxvaerd, and Dyre}}]{moleculeshidden}
\bibinfo{author}{\bibfnamefont{T.~B.} \bibnamefont{Schr{\o}der}},
  \bibinfo{author}{\bibfnamefont{U.~R.} \bibnamefont{Pedersen}},
  \bibinfo{author}{\bibfnamefont{N.~P.} \bibnamefont{Bailey}},
  \bibinfo{author}{\bibfnamefont{S.}~\bibnamefont{Toxvaerd}}, \bibnamefont{and}
  \bibinfo{author}{\bibfnamefont{J.~C.} \bibnamefont{Dyre}},
  \bibinfo{journal}{Phys. Rev. E} \textbf{\bibinfo{volume}{80}},
  \bibinfo{pages}{041502} (\bibinfo{year}{2009}).

\bibitem[{\citenamefont{Schoeffel et~al.}(2012)\citenamefont{Schoeffel,
  Brodie-Linder, Audonnetx, and Alba-Simionesco}}]{schoeffel2012}
\bibinfo{author}{\bibfnamefont{M.}~\bibnamefont{Schoeffel}},
  \bibinfo{author}{\bibfnamefont{N.}~\bibnamefont{Brodie-Linder}},
  \bibinfo{author}{\bibfnamefont{F.}~\bibnamefont{Audonnetx}},
  \bibnamefont{and}
  \bibinfo{author}{\bibfnamefont{C.}~\bibnamefont{Alba-Simionesco}},
  \bibinfo{journal}{J. Mater. Chem.} \textbf{\bibinfo{volume}{22}},
  \bibinfo{pages}{557} (\bibinfo{year}{2012}).

\bibitem[{\citenamefont{Krekelberg et~al.}(2011)\citenamefont{Krekelberg, Shen,
  Errington, and Truskett}}]{krekelberg2011}
\bibinfo{author}{\bibfnamefont{W.~P.} \bibnamefont{Krekelberg}},
  \bibinfo{author}{\bibfnamefont{V.~K.} \bibnamefont{Shen}},
  \bibinfo{author}{\bibfnamefont{J.~R.} \bibnamefont{Errington}},
  \bibnamefont{and} \bibinfo{author}{\bibfnamefont{T.~M.}
  \bibnamefont{Truskett}}, \bibinfo{journal}{J. Chem. Phys.}
  \textbf{\bibinfo{volume}{135}}, \bibinfo{pages}{154502}
  (\bibinfo{year}{2011}).

\bibitem[{\citenamefont{Debenedetti and Stillinger}(2001)}]{debenedetti2001}
\bibinfo{author}{\bibfnamefont{P.~G.} \bibnamefont{Debenedetti}}
  \bibnamefont{and} \bibinfo{author}{\bibfnamefont{F.~H.}
  \bibnamefont{Stillinger}}, \bibinfo{journal}{Nature}
  \textbf{\bibinfo{volume}{410}}, \bibinfo{pages}{259} (\bibinfo{year}{2001}).

\bibitem[{\citenamefont{Dosseh et~al.}(2006)\citenamefont{Dosseh, Quellec,
  Brodie-linder, Alba-simionesco, Haeussler, and Levitz}}]{dosseh2006}
\bibinfo{author}{\bibfnamefont{G.}~\bibnamefont{Dosseh}},
  \bibinfo{author}{\bibfnamefont{C.~L.} \bibnamefont{Quellec}},
  \bibinfo{author}{\bibfnamefont{N.}~\bibnamefont{Brodie-linder}},
  \bibinfo{author}{\bibfnamefont{C.}~\bibnamefont{Alba-simionesco}},
  \bibinfo{author}{\bibfnamefont{W.}~\bibnamefont{Haeussler}},
  \bibnamefont{and} \bibinfo{author}{\bibfnamefont{P.}~\bibnamefont{Levitz}},
  \bibinfo{journal}{J. Non-Cryst. Solids} \textbf{\bibinfo{volume}{352}},
  \bibinfo{pages}{4964} (\bibinfo{year}{2006}).

\bibitem[{\citenamefont{Cohen and Grest}(1979)}]{cohen1979}
\bibinfo{author}{\bibfnamefont{M.~H.} \bibnamefont{Cohen}} \bibnamefont{and}
  \bibinfo{author}{\bibfnamefont{G.~S.} \bibnamefont{Grest}},
  \bibinfo{journal}{Phys. Rev. B} \textbf{\bibinfo{volume}{20}},
  \bibinfo{pages}{1077} (\bibinfo{year}{1979}).

\bibitem[{\citenamefont{Rosenfeld}(1977)}]{rosenfeld1}
\bibinfo{author}{\bibfnamefont{Y.}~\bibnamefont{Rosenfeld}},
  \bibinfo{journal}{Phys. Rev. A} \textbf{\bibinfo{volume}{15}},
  \bibinfo{pages}{2545} (\bibinfo{year}{1977}).

\bibitem[{\citenamefont{Gnan et~al.}(2009)\citenamefont{Gnan, Schr{\o}der,
  Pedersen, Bailey, and Dyre}}]{paper4}
\bibinfo{author}{\bibfnamefont{N.}~\bibnamefont{Gnan}},
  \bibinfo{author}{\bibfnamefont{T.~B.} \bibnamefont{Schr{\o}der}},
  \bibinfo{author}{\bibfnamefont{U.~R.} \bibnamefont{Pedersen}},
  \bibinfo{author}{\bibfnamefont{N.~P.} \bibnamefont{Bailey}},
  \bibnamefont{and} \bibinfo{author}{\bibfnamefont{J.~C.} \bibnamefont{Dyre}},
  \bibinfo{journal}{J. Chem. Phys.} \textbf{\bibinfo{volume}{131}},
  \bibinfo{pages}{234504} (\bibinfo{year}{2009}).

\bibitem[{\citenamefont{Bailey et~al.}(2008)\citenamefont{Bailey, Pedersen,
  Gnan, Schr{\o}der, and Dyre}}]{paper1}
\bibinfo{author}{\bibfnamefont{N.~P.} \bibnamefont{Bailey}},
  \bibinfo{author}{\bibfnamefont{U.~R.} \bibnamefont{Pedersen}},
  \bibinfo{author}{\bibfnamefont{N.}~\bibnamefont{Gnan}},
  \bibinfo{author}{\bibfnamefont{T.~B.} \bibnamefont{Schr{\o}der}},
  \bibnamefont{and} \bibinfo{author}{\bibfnamefont{J.~C.} \bibnamefont{Dyre}},
  \bibinfo{journal}{J. Chem. Phys.} \textbf{\bibinfo{volume}{129}},
  \bibinfo{pages}{184507} (\bibinfo{year}{2008}).

\bibitem[{\citenamefont{Ingebrigtsen
  et~al.}(2012{\natexlab{a}})\citenamefont{Ingebrigtsen, Schr{\o}der, and
  Dyre}}]{prx}
\bibinfo{author}{\bibfnamefont{T.~S.} \bibnamefont{Ingebrigtsen}},
  \bibinfo{author}{\bibfnamefont{T.~B.} \bibnamefont{Schr{\o}der}},
  \bibnamefont{and} \bibinfo{author}{\bibfnamefont{J.~C.} \bibnamefont{Dyre}},
  \bibinfo{journal}{Phys. Rev. X} \textbf{\bibinfo{volume}{2}},
  \bibinfo{pages}{011011} (\bibinfo{year}{2012}{\natexlab{a}}).

\bibitem[{\citenamefont{Gundermann et~al.}(2011)\citenamefont{Gundermann,
  Pedersen, Hecksher, Bailey, Jakobsen, Christensen, Olsen, Schr{\o}der,
  Fragiadakis, Casalini et~al.}}]{gammagamma}
\bibinfo{author}{\bibfnamefont{D.}~\bibnamefont{Gundermann}},
  \bibinfo{author}{\bibfnamefont{U.~R.} \bibnamefont{Pedersen}},
  \bibinfo{author}{\bibfnamefont{T.}~\bibnamefont{Hecksher}},
  \bibinfo{author}{\bibfnamefont{N.~P.} \bibnamefont{Bailey}},
  \bibinfo{author}{\bibfnamefont{B.}~\bibnamefont{Jakobsen}},
  \bibinfo{author}{\bibfnamefont{T.}~\bibnamefont{Christensen}},
  \bibinfo{author}{\bibfnamefont{N.~B.} \bibnamefont{Olsen}},
  \bibinfo{author}{\bibfnamefont{T.~B.} \bibnamefont{Schr{\o}der}},
  \bibinfo{author}{\bibfnamefont{D.}~\bibnamefont{Fragiadakis}},
  \bibinfo{author}{\bibfnamefont{R.}~\bibnamefont{Casalini}},
  \bibnamefont{et~al.}, \bibinfo{journal}{Nat. Phys.}
  \textbf{\bibinfo{volume}{7}}, \bibinfo{pages}{816} (\bibinfo{year}{2011}).

\bibitem[{\citenamefont{Ingebrigtsen
  et~al.}(2012{\natexlab{b}})\citenamefont{Ingebrigtsen, Schr{\o}der, and
  Dyre}}]{moleculesisomorphs}
\bibinfo{author}{\bibfnamefont{T.~S.} \bibnamefont{Ingebrigtsen}},
  \bibinfo{author}{\bibfnamefont{T.~B.} \bibnamefont{Schr{\o}der}},
  \bibnamefont{and} \bibinfo{author}{\bibfnamefont{J.~C.} \bibnamefont{Dyre}},
  \bibinfo{journal}{J. Phys. Chem. B} \textbf{\bibinfo{volume}{116}},
  \bibinfo{pages}{1018} (\bibinfo{year}{2012}{\natexlab{b}}).

\bibitem[{\citenamefont{Ingebrigtsen et~al.}(2013)\citenamefont{Ingebrigtsen,
  Veldhorst, Schr{\"o}der, and Dyre}}]{RTingebrigtsen}
\bibinfo{author}{\bibfnamefont{T.~S.} \bibnamefont{Ingebrigtsen}},
  \bibinfo{author}{\bibfnamefont{A.~A.} \bibnamefont{Veldhorst}},
  \bibinfo{author}{\bibfnamefont{T.~B.} \bibnamefont{Schr{\"o}der}},
  \bibnamefont{and} \bibinfo{author}{\bibfnamefont{J.~C.} \bibnamefont{Dyre}},
  \bibinfo{journal}{J. Chem. Phys.} \textbf{\bibinfo{volume}{139}},
  \bibinfo{pages}{171101} (\bibinfo{year}{2013}).

\bibitem[{\citenamefont{Shintani and Tanaka}(2006)}]{shintani2006}
\bibinfo{author}{\bibfnamefont{H.}~\bibnamefont{Shintani}} \bibnamefont{and}
  \bibinfo{author}{\bibfnamefont{H.}~\bibnamefont{Tanaka}},
  \bibinfo{journal}{Nature} \textbf{\bibinfo{volume}{2}}, \bibinfo{pages}{200}
  (\bibinfo{year}{2006}).

\bibitem[{\citenamefont{Kawasaki et~al.}(2007)\citenamefont{Kawasaki, Araki,
  and Tanaka}}]{kawasaki2007}
\bibinfo{author}{\bibfnamefont{T.}~\bibnamefont{Kawasaki}},
  \bibinfo{author}{\bibfnamefont{T.}~\bibnamefont{Araki}}, \bibnamefont{and}
  \bibinfo{author}{\bibfnamefont{H.}~\bibnamefont{Tanaka}},
  \bibinfo{journal}{Phys. Rev. Lett.} \textbf{\bibinfo{volume}{99}},
  \bibinfo{pages}{215701} (\bibinfo{year}{2007}).

\bibitem[{\citenamefont{Leocmach and Tanaka}(2012)}]{leocmach2012}
\bibinfo{author}{\bibfnamefont{M.}~\bibnamefont{Leocmach}} \bibnamefont{and}
  \bibinfo{author}{\bibfnamefont{H.}~\bibnamefont{Tanaka}},
  \bibinfo{journal}{Nat. Commun.} \textbf{\bibinfo{volume}{3}},
  \bibinfo{pages}{974} (\bibinfo{year}{2012}).

\bibitem[{\citenamefont{Krakoviack}(2005)}]{krakoviack2005}
\bibinfo{author}{\bibfnamefont{V.}~\bibnamefont{Krakoviack}},
  \bibinfo{journal}{Phys. Rev. Lett.} \textbf{\bibinfo{volume}{94}},
  \bibinfo{pages}{065703} (\bibinfo{year}{2005}).

\bibitem[{\citenamefont{Biroli et~al.}(2006)\citenamefont{Biroli, Bouchaud,
  Miyazaki, and Reichman}}]{biroli2006}
\bibinfo{author}{\bibfnamefont{G.}~\bibnamefont{Biroli}},
  \bibinfo{author}{\bibfnamefont{J.-P.} \bibnamefont{Bouchaud}},
  \bibinfo{author}{\bibfnamefont{K.}~\bibnamefont{Miyazaki}}, \bibnamefont{and}
  \bibinfo{author}{\bibfnamefont{D.~R.} \bibnamefont{Reichman}},
  \bibinfo{journal}{Phys. Rev. Lett.} \textbf{\bibinfo{volume}{97}},
  \bibinfo{pages}{195701} (\bibinfo{year}{2006}).

\bibitem[{\citenamefont{Lang et~al.}(2010)\citenamefont{Lang, Botan, Oettel,
  Hajnal, Franosch, and Schilling}}]{lang2010}
\bibinfo{author}{\bibfnamefont{S.}~\bibnamefont{Lang}},
  \bibinfo{author}{\bibfnamefont{V.}~\bibnamefont{Botan}},
  \bibinfo{author}{\bibfnamefont{M.}~\bibnamefont{Oettel}},
  \bibinfo{author}{\bibfnamefont{D.}~\bibnamefont{Hajnal}},
  \bibinfo{author}{\bibfnamefont{T.}~\bibnamefont{Franosch}}, \bibnamefont{and}
  \bibinfo{author}{\bibfnamefont{R.}~\bibnamefont{Schilling}},
  \bibinfo{journal}{Phys. Rev. Lett.} \textbf{\bibinfo{volume}{105}},
  \bibinfo{pages}{125701} (\bibinfo{year}{2010}).

\bibitem[{\citenamefont{Lang et~al.}(2012)\citenamefont{Lang, Schilling,
  Krakoviack, and Franosch}}]{lang2012}
\bibinfo{author}{\bibfnamefont{S.}~\bibnamefont{Lang}},
  \bibinfo{author}{\bibfnamefont{R.}~\bibnamefont{Schilling}},
  \bibinfo{author}{\bibfnamefont{V.}~\bibnamefont{Krakoviack}},
  \bibnamefont{and} \bibinfo{author}{\bibfnamefont{T.}~\bibnamefont{Franosch}},
  \bibinfo{journal}{Phys. Rev. E} \textbf{\bibinfo{volume}{86}},
  \bibinfo{pages}{021502} (\bibinfo{year}{2012}).

\bibitem[{\citenamefont{van Ketel et~al.}(2005)\citenamefont{van Ketel, Das,
  and Frenkel}}]{ketel2005}
\bibinfo{author}{\bibfnamefont{W.}~\bibnamefont{van Ketel}},
  \bibinfo{author}{\bibfnamefont{C.}~\bibnamefont{Das}}, \bibnamefont{and}
  \bibinfo{author}{\bibfnamefont{D.}~\bibnamefont{Frenkel}},
  \bibinfo{journal}{Phys. Rev. Lett.} \textbf{\bibinfo{volume}{94}},
  \bibinfo{pages}{135703} (\bibinfo{year}{2005}).

\bibitem[{\citenamefont{Adam and Gibbs}(1965)}]{adam1965}
\bibinfo{author}{\bibfnamefont{G.}~\bibnamefont{Adam}} \bibnamefont{and}
  \bibinfo{author}{\bibfnamefont{J.~H.} \bibnamefont{Gibbs}},
  \bibinfo{journal}{J. Chem. Phys.} \textbf{\bibinfo{volume}{43}},
  \bibinfo{pages}{139} (\bibinfo{year}{1965}).

\bibitem[{\citenamefont{Nos\'{e}}(1984)}]{nose}
\bibinfo{author}{\bibfnamefont{S.}~\bibnamefont{Nos\'{e}}},
  \bibinfo{journal}{J. Chem. Phys.} \textbf{\bibinfo{volume}{81}},
  \bibinfo{pages}{511} (\bibinfo{year}{1984}).

\bibitem[{\citenamefont{Ingebrigtsen
  et~al.}(2012{\natexlab{c}})\citenamefont{Ingebrigtsen, B{\o}hling,
  Schr{\o}der, and Dyre}}]{thermoscl}
\bibinfo{author}{\bibfnamefont{T.~S.} \bibnamefont{Ingebrigtsen}},
  \bibinfo{author}{\bibfnamefont{L.}~\bibnamefont{B{\o}hling}},
  \bibinfo{author}{\bibfnamefont{T.~B.} \bibnamefont{Schr{\o}der}},
  \bibnamefont{and} \bibinfo{author}{\bibfnamefont{J.~C.} \bibnamefont{Dyre}},
  \bibinfo{journal}{J. Chem. Phys.} \textbf{\bibinfo{volume}{136}},
  \bibinfo{pages}{061102} (\bibinfo{year}{2012}{\natexlab{c}}).

\end{thebibliography}

\end{document}